\documentclass[12pt]{article}
\usepackage{graphicx}
\usepackage{amsmath}
\usepackage{amssymb}
\usepackage{amsbsy}
\usepackage{bm}   % for bold math fonts, especially greek
\usepackage{framed}  % for frame boxes
\numberwithin{equation}{section}

\newcommand{\bit}{\begin{itemize}}
\newcommand{\eit}{\end{itemize}}

\def\benu{\begin{enumerate}}
\def\eenu{\end{enumerate}}
\def\noi{\noindent}
\def\btab{\begin{tabbing}}
\def\etab{\end{tabbing}}

\def\bit{\begin{itemize}}
\def\eit{\end{itemize}}
\def\beq{\begin{equation}}
\def\eeq{\end{equation}}
\def\bec{\begin{center}}
\def\eec{\end{center}}
\def\btable{\begin{tabular}}
\def\etable{\end{tabular}}
\def\beqr{\begin{eqnarray}}
\def\eeqr{\end{eqnarray}}
\def\rarw{\rightarrow}

\def\om{\omega}

\def\eps{\epsilon}

\def\al{\alpha}
\def\bt{\beta}

\def\dl{\delta}
\def\Dl{\Delta}
\def\sg{\sigma}

\def\rarw{\rightarrow}

\def\half{\frac{1}{2}}
\def\qrtr{\frac{1}{4}}

\def\btab{\begin{tabbing}}
\def\etab{\end{tabbing}}
\def\beqrs{\begin{eqnarray*}}
\def\eeqrs{\end{eqnarray*}}
\def\noi{\noindent}
\def\lan{\langle}
\def\ran{\rangle}
\def\bfig{\begin{figure}}
\def\efig{\end{figure}}
\def\fr{\frac}

 \fontfamily{ptm}\selectfont
 \usepackage{mathptmx}           % this gives slanted Times symbols for math

\title{\bf Nonlinear theory of  transverse beam echoes}
\author{Tanaji Sen \footnote{tsen@fnal.gov} \\ Accelerator Physics Center, FNAL, Batavia, IL 60510 \\
Yuan Shen Li \footnote{Present address: Dept. of Physics, University of Chicago, Chicago, IL 60637} \\
Carleton College, Northfield, MN 55057
}
\date{}

\begin{document}

\maketitle

\begin{abstract}
Transverse beam echoes can be excited with a single dipole kick followed by a single quadrupole kick. They have been used
to measure diffusion in hadron beams and have other diagnostic capabilities.  Here 
we develop  theories of the transverse echo nonlinear in both the dipole and quadrupole kick strengths. 
The theories predict the maximum echo amplitudes and the optimum strength parameters. We find that the echo amplitude
increases with smaller beam emittance and the asymptotic echo amplitude can exceed half the initial dipole kick amplitude. 
We show that multiple echoes can be observed provided the dipole kick is large enough. The spectrum of the echo pulse
can be used to determine the nonlinear detuning parameter with
small amplitude dipole kicks.  Simulations are performed to check the theoretical predictions. In the useful ranges  of dipole and
quadrupole strengths, they are shown to be in reasonable agreement. 
\end{abstract}

\maketitle

%\tableofcontents

\section{Introduction}

Echoes are ubiquitous phenomena in physics. Spin echoes were discovered by Hahn \cite{Hahn} and since
then, spin echoes have evolved into use as sophisticated diagnostic tools in magnetic resonance imaging
\cite{Dale}. Photon echoes were observed from a ruby crystal after excitation by a sequence of two
laser pulses, each about 0.1$\mu$s long \cite{Kurnit}. Later, 
plasma wave echoes were predicted and then observed in a plasma excited by two rf pulses \cite{Gould, Malmberg}. 
A system of ultra-cold atoms confined within an optical trap exhibited echoes when excited by a sequence of microwave pulses
\cite{Andersen}. 
About a decade ago, fluid echoes were observed in a magnetized electron plasma \cite{Yu}. 
More recently, so called fractional echoes were observed in a CO$_2$ gas excited by two femtosecond laser pulses \cite{Karras}. 
Echoes were first introduced into accelerator physics more than two decades ago \cite{Stupakov, Stup_Kauf}. This was followed by the observation of
longitudinal echoes in unbunched beams first at the Fermilab Antiproton Accumulator \cite{Fermi_AA} and later at the SPS
 \cite{CERN_SPS}. Transverse echoes were seen at the SPS \cite{Arduini}, but more detailed studies with transverse
bunched beam echoes were performed at RHIC \cite{Fischer_2005}. A detailed analysis of these experiments to extract
diffusion coefficients was recently reported in \cite{Sen_2017}. 

In all echo phenomena, the system (atoms, plasma, particle beam etc.) is first acted on by a pulsed excitation (e.g. a dipole 
kick on a beam) that excites a coherent response which then decoheres due to phase mixing. However the information in the 
macroscopic observables (coordinate moments for a particle beam) is not lost, but can be retrieved by the application of a second
pulsed excitation (e.g. a quadrupole kick). Some time after the response to the second excitation has disappeared, 
a coherent response, called the echo, reappears. The strength of the echo signal in a beam depends on the beam parameters
and on the strengths of the kicks from the magnets. The echo response is exquisitely sensitive to the presence of beam
diffusion. This sensitivity  simultaneously presents both opportunities and challenges. The short time scale over which beam echoes can be
measured (typically within a few thousand turns in an accelerator ring) implies that diffusion can be measured very quickly 
compared to the conventional method of using movable collimators e.g. \cite{Stancari}, which can take hours. However, 
the echo signal can also be destroyed by strong diffusion. It is therefore necessary to understand how to 
maximize the echo response by appropriate choices of beam parameters and excitation strengths. 

In this paper we develop  a  theory of echoes in one degree of freedom with nonlinear dependence on dipole and quadrupole 
strengths, with the goal of maximizing the echo signal. A nonlinear theory had been developed earlier in \cite{Stup_Kauf}. Here 
we follow a different approach, the method as described in \cite{Chao} where it was restricted to a linear theory. Our results are more general than those in \cite{Stup_Kauf}, but reduce to them in limiting cases. 
In Section II, we develop a theory (labeled QT) that is
linear in the dipole kick, but nonlinear in the quadrupole kick strength. This is followed in Section III with a simplified echo theory 
(labeled DQT)  that is nonlinear in both dipole and quadrupole kick strengths; a more complete theory is described in 
Appendix A.
Section IV discusses  simulations performed to check the
theoretical results. Section V shows how the spectrum of the echo pulse can be used to extract the detuning parameter and we
end in Section VI with our conclusions.

\setcounter{equation}{0}
\section{Nonlinear theory of quadrupole kicks} \label{sec: NL_quad}

The simplest way to generate a transverse beam echo is to apply a short pulse dipole kick, usually done with an injection
kicker, to a beam in an accelerator ring with nonlinear elements so that the betatron tune is amplitude dependent. 
The centroid motion decoheres due to the tune spread \cite{Meller} and at some time $\tau$ after the dipole kick, the beam is
excited with a 
short pulse quadrupole kick. For simplicity we will consider a single turn quadrupole kick, although this is strictly not necessary 
and this kick could last a few turns. Following the quadrupole kick, the decoherence starts to partially reverse and at time 
2$\tau$ after the dipole kick, the first echo appears. Depending on beam parameters and kick strengths, multiple echoes can appear at times 
$4\tau$, $6\tau$ etc. 

The echo amplitude depends on several parameters, especially the dipole and quadrupole  kick strengths.
Our approach will be to develop an Eulerian theory by following the flow of the density distribution, similar to the development in 
\cite{Chao} where both kicks were treated in linearized approximations. In this section, we develop a theory (labeled QT)
that is linear in the dipole strength but nonlinear in the quadrupole strength. 
We will compare our results with those from an
alternative method of following the particle's phase space motion that had been developed earlier \cite{Stup_Kauf}.
As mentioned in the Introduction, the treatment here is for motion in one transverse degree of freedom, so the effects of 
transverse coupling as well as coupling to the effects of synchrotron oscillations and energy spread are ignored here. 
We also do not consider here how diffusion reduces the echo amplitudes or the impact of collective effects at high
intensity. These are important effects which will be considered
elsewhere. 

We start with the usual definitions  of the phase space variables in position and momentum $(x, p)$ and the corresponding
action and angle variables $(J,\phi)$
\beqr
 x & = &  \sqrt{2 \bt J} \cos\phi, \;\;\;\;  p = \bt x' + \al x = - \sqrt{2 \bt  J} \sin\phi \\
J & =  & \frac{1}{2\bt}[x^2 + p^2], \;\;\;\; \phi = {\rm Arctan}(\frac{-p}{x}) 
\eeqr
We will assume that the nonlinear motion of the particles can be modeled by an action dependent betatron frequency
and for simplicity we assume the form
\beq
\om(J) = \om_{\bt} + \om' J
\eeq
where $\om_{\bt}$ is the bare angular betatron frequency, and $\om'$ is the frequency slope which is determined by the
lattice nonlinearities. This model therefore assumes that the effects of nearby resonances are negligible. 
We assume that the initial particle distribution is a Gaussian in $(x,p)$ or equivalently an exponential in the action 
\beq
\psi_0(J) = \frac{1}{2\pi \eps_0} \exp[-\frac{J}{\eps_0}]
\eeq
with initial emittance $\eps_0$. 
At time $t=0$, an impulsive single turn dipole kick $\Dl p = \bt_K \Dl x' = \bt_K \theta$ changes the distribution function (DF) to 
$\psi_1(J, \phi) = \psi_0(x,p - \bt_K\theta)$ where $\bt_K$ is the beta function at the dipole and $\theta$ is the kick angle. To 
first order in the dipole kick, we have
\beq
\psi_1(J, \phi) = \psi_0(J) + \bt_K \theta \psi_0'(J) \sqrt{\frac{2J}{\bt}}\sin\phi
\eeq
Following the dipole kick, the action remains constant while the angle $\phi$ evolves by a free betatron rotation. 
Hence, at time $t$ after the dipole kick, the DF is 
\beq
\psi_2(J,\phi,t) = \psi_0(J) + \bt_K \theta \psi_0'(J) \sqrt{\frac{2J}{\bt}}\sin(\phi - \om(J)t)
\eeq
Just before the quadrupole kick at time $\tau$, the DF is $\psi_3(J,\phi,\tau)  = \psi_2(J,\phi,t=\tau) $. 
The first term $\psi_0(J)$ in the perturbed DF does not contribute to the dipole moment, and it will be dropped in the rest of this 
section.
The quadrupole kick $\Dl p = - qx$ changes the distribution to $\psi_4(x,p,\tau) = \psi_3(x, p + q x,\tau)$. 
Here $q = \bt_Q/f$ is the dimensionless quadrupole strength parameter, with $\bt_Q$ the beta function at the quadrupole
and $f$ the focal length of this quadrupole. In practical applications $q \ll 1$ and we will assume this to be true in the
development here. 

Due to this quadrupole kick, the action and angle arguments of the density distribution change to
\beqr
J & \rarw & \frac{1}{2\bt}[x^2 + (p+qx)^2]   \equiv J[1 + A(q,\phi)], \;\;\;  A(q,\phi) =  (-q \sin 2\phi + q^2 \cos^2\phi) 
\label{eq: Jtranform_A} \\
\phi & \rarw &  {\rm Arctan}(-\frac{p + qx}{x}) = {\rm Arctan}(\tan\phi - q) 
\eeqr

To proceed, we have to approximate the form of the transformed angle variable. A Taylor expansion shows that
\beq
  {\rm Arctan}(\tan\phi - q) = \phi - q \cos^2\phi - \qrtr q^2(\sin 2\phi + \half \sin 4\phi)  + O(q^3) 
\eeq
For reasons of simpliciity, we will keep terms to $O(q)$ in this expansion. 
For self- consistency,  we consider $A(q,\phi)$ to the same order and approximate $A(q,\phi) \approx - q \sin 2\phi$. 
While the Jacobian of the exact transformation has a determinant of one, the approximate transforms has the 
determinant $= 1 +  O(q^2)$.

The DF right after the quadrupole kick with the approximation above is given by,
\beqr
\psi_4(J, \phi,\tau) & =  & \bt_K\theta \psi_0'(J[1 - q\sin 2\phi]) \sqrt{\frac{2J(1 - q\sin 2\phi)}{\bt}}
\sin\left[\phi_{-\tau} - q\cos^2\phi \right]  \\
\phi_{-\tau}& =  & \phi - \om(J[1- q\sin 2\phi]) \tau 
\eeqr
Following the quadrupole kick, the DF at time $t$ (from the instant of the dipole kick) is
\beq
\psi_5(J, \phi,t) = \psi_4(J, \phi_{-\Dl \phi}), \;\;\;\;  \phi_{-\Dl \phi} \equiv \phi -  \Dl\phi, \;\;\; \Dl\phi = \om(J)(t - \tau)
\eeq
We note that as defined here, $\Dl\phi$ depends on the action $J$ but is independent of the angle $\phi$. 
Under the change $\phi \rarw \phi_{-\Dl \phi}$, the angle variable $\phi_{-\tau}$ transforms as
$ \phi_{-\tau} \rarw   \phi_{- \Dl\phi} -  \tau \om + q\tau\om' J \sin 2\phi_{-\Dl \phi}$. 
The dipole moment at time $t$ is
\beqr
\lan x \ran(t) & = &  \sqrt{2\bt}\int dJ \int d\phi \sqrt{J}\cos\phi \psi_5(J,\phi,t) \nonumber \\
& = & 2 \bt_K \theta \int dJ \int d\phi \sqrt{J}\cos\phi  \psi_0'(J[1- q\sin 2\phi_{-\Dl\phi})]) 
\sqrt{J[1- q\sin 2\phi_{-\Dl\phi})]}  \nonumber \\
& & \times \sin\left(\phi_{-\Dl\phi} - \half q(1+ \cos 2\phi_{-\Dl\phi}) - \tau\om  +
q\tau\om' J \sin 2\phi_{-\Dl \phi}\right)
\label{eq: dipmom_theta1}
\eeqr
We proceed by simplifying the trigonometric terms in the argument of the first sine function in the last line above
\beqr
- \half \cos  2\phi_{-\Dl\phi} + \tau\om' J \sin  2\phi_{-\Dl\phi} & = & \sqrt{(\tau\om' J)^2 + \qrtr}
\sin[2\phi_{-\Dl\phi} - {\rm Arctan}(\frac{1}{2\tau \om' J})]  \nonumber \\ 
 & \approx & \tau\om' J \sin 2\phi_{-\Dl\phi}   \label{eq: approx_tauD}
\eeqr
where the last approximation follows by noting that the decoherence time $\tau_D \simeq \om' \eps_0$ is much shorter 
than the delay time $\tau$, hence $\tau \om' \eps_0 \simeq \tau/\tau_D \gg 1$. Next, we 
expand the square root to first order in $q$ as 
$\sqrt{[1 - q\sin 2\phi_{-\Dl\phi})]}  \approx \left[ 1 -\half q \sin 2\phi_{-\Dl \phi} \right]$.

Hence we can write
\beq
\lan x(t)\ran = -\frac{\bt_K \theta}{2\pi \eps_0^2} \int  J \exp[-\frac{J}{\eps_0}]
\left\{ S_1 - S_2 + S_3 - S_4  \right\} dJ \equiv  T_1 - T_2 + T_3 - T_4    \label{eq: def_T_i}
\eeq
The terms $S_i$ are obtained after integrating over $\phi$ and are given by
\beqr
S_1 & = &- 2\pi  \; {\rm Im}\left\{\exp[i( \Dl\phi - \tau \om - \half q )] J_{1}(q\tau\om'  J) \right\} \\
S_2 & = & 2\pi  {\rm Im}\left\{ \exp[i (\Dl\phi + \tau\om + \half q )] J_0(q\tau\om'  J) \right\} \\
S_3 & = & -\frac{\pi}{2} q{\rm Re}\left\{ \exp[i(-\Dl\phi + \tau\om + \half q ) ] J_0(q\tau\om'  J)  - \exp[i(\Dl\phi - \tau\om - \half q )] J_{2}(q\tau\om'  J) \right\}  \nonumber \\
\mbox{} \\
S_4 & = &  \; \frac{\pi}{2} {\rm Re} \left\{ \exp[-i(\Dl\phi + \tau\om + \half q )] J_{-1}(q\tau\om'  J) + 
\exp[i(\Dl\phi + \tau\om)] J_{1}(q\tau\om'  J) \right\} 
\eeqr
where the integrals over $\phi$  were done by  first expanding into Bessel functions and using
\beqr
\int d\phi \exp[i m\phi] \exp[i a \sin (2\phi - 2\Dl\phi)] & = & \int d\phi \exp[i m\phi]  \sum_k  J_k(a) \exp[i k (2\phi - 2\Dl\phi)]  \nonumber \\
& = & 2\pi J_{-m/2}(a) \exp[i m \Dl\phi] 
\eeqr
We clarify that $J$ denotes the action while $J_n$ with a subscript $n$ will denote the Bessel function. 

To integrate over the action $J$, we introduce the dimensionless integration variable $z = J/\eps_0$ and define
the following dimensionless parameters that are independent of the action,
\beqr
 \Phi & = & \om_{\bt} (t - 2\tau) , \;\;\; \xi(t) =  (t - 2\tau)\om' \eps_0 , \;\;\;  Q  =   q \tau \om' \eps_0   \\
 a_1 & = & 1 - i\xi  , \;\;\;\; a_2  =  1  - i  \om' t \eps_0 
\eeqr
It follows that the terms $T_i$, obtained by integrating over $J$ in Eq.~(\ref{eq: def_T_i}), are 
\beqr
T_1 & = &  \bt_K \theta  \; {\rm Im}\left\{ \exp[i(\Phi - \half q)] H_{1, 1}(a_1, Q)  \right\} \\
T_2 & = &  -\bt_K \theta  \; {\rm Im}\left\{ \exp[i(\om_{\bt} t + \half q)] H_{1, 0}(a_2, Q) \right\} \\
T_3 & = &   -\qrtr \bt_K \theta q  \; {\rm Re}\left\{ \exp[-i(\Phi - \half q)] H_{1, 0}(a_1^*, Q)  
 - \exp[i(\Phi - \half q)]    H_{1, 2}(a_1, Q) \right\} \\
T_4 & = &  \qrtr \bt_K \theta \eps_0^2 q \; {\rm Re}\left\{ \exp[-i(\om_{\bt} t + \half q)] H_{1, 1}(a_2^* , Q)
 + \exp[i(\om_{\bt} t + \half q)]   H_{1, 1}(a_2, Q) \right\}  \nonumber \\
\mbox{}
\eeqr
where $a_1^*$ is the complex conjugate of $a_1$ and the functions $H_{m, n}(a, Q)$ are defined as
\beq
H_{m, n}(a, Q) = \int_0^{\infty}dz \; z^m \exp[- a z] J_n(Q z)
\eeq
Consider only the terms with phases that depend on $\Phi$ rather than on $\om_{\bt}t$. These phase
terms will vanish around the time of the echo at $t=2\tau$ and the terms $T_1, T_3$ will  be the
dominant terms to determine the echo amplitude.  
\beqr
T_1 & = &  \bt_K \theta  \; {\rm Im}\left\{ \exp[i(\Phi - \half q)] \frac{Q}{(a_1^2 + Q^2)^{3/2}} \right\} \\
T_3 & = & -\qrtr \bt_K \theta q \; {\rm Re}\left\{ \exp[-i(\Phi - \half q)]  \frac{a_1^*}{((a_1^*)^2 + Q^2)^{3/2}} \right.
 \nonumber \\
& & \left.  - \exp[i(\Phi - \half q)]    \fr{2 (a_1^2 + Q^2)^{3/2} - a_1(2a_1^2 +3 Q^2)}{Q^2 (a_1^2 + Q^2)^{3/2}} \right\} 
\eeqr
In order to simplify the evaluation of these terms, we introduce
the amplitude functions $A_0(t; \tau, q), A_1(t; \tau, q)$, the phase functions $\Theta(t; \tau, q), \Theta_1(t; \tau, q)$ and two 
other terms $a_{3C}, a_{3S}$ as follows
\beqr
(a_1^2 + Q^2)^{3/2} & \equiv & A \exp[-i 3\Theta],  \; \; \; \;  a_1  \equiv   A_1(t; \tau, q)\exp[i\Theta_1]  \\ 
A_0(t; \tau, q)  & =  &  [(1 - \xi ^2 + Q^2)^2  + 4 \xi^2]^{3/4} , \;\;\;
 \Theta   =    {\rm Arctan}[\frac{\xi }{1 - \xi ^2 + Q^2}]  \label{eq: Theta} \\
A_1 & = &  [ 1 + \xi ^2]^{1/2},  \;\;\; \Theta_1 = {\rm Arctan}[\xi ]  \\
 a_{3C} & = & -\half q\left( A_1\cos(3\Theta + \Theta_1) - \fr{A_0}{Q^2} + \fr{A_0^{2/3}A_1}{Q^2}\cos(\Theta+ \Theta_1) \right) \\
 a_{3S} & = &  -\half q\left(A_1\sin(3\Theta + \Theta_1) + \fr{A_0^{2/3}A_1}{Q^2}\sin(\Theta+ \Theta_1) \right) 
\eeqr
In terms of these amplitudes and phases, the functions $T_1, T_3$ simplify to
\beq
T_1  =    \frac{\bt_K \theta Q }{A_0}\sin[\Phi  - \half q + 3\Theta] , \;\;\;
T_3  =  - \fr{\bt_K \theta}{A_0}\left[ a_{3C}\cos(\Phi - \half q) - a_{3S}\sin (\Phi - \half q) \right]
\eeq
Keeping these two dominant terms at large times, we can write the time dependent echo in terms of an amplitude and phase as
\beqr
\lan x(t) \ran & = & T_1 +  T_3 =   \bt_K \theta A_{1,3}\sin(\Phi(t) + \Theta_{1,3}(t) - \half q)   \label{eq: amp_nonlinq_gen} \\
A_{1, 3} & = & \frac{1}{A_0}\left[ (Q \cos 3\Theta + a_{3S})^2 + (Q \sin 3\Theta - a_{3C})^2
\right]^{1/2} \label{eq: A13} \\
\Theta_{1, 3} & \equiv & {\rm Arctan}\left[ \frac{Q \sin 3\Theta - a_{3C}}{Q \cos 3\Theta + a_{3S}}\right]
\eeqr
We consider various limiting forms of this general form of the echo (in the linearized dipole kick approximation of this 
section) below. 

Of the two terms, $T_1$ has the dominant contribution to the echo amplitude. Keeping only this term, the time
dependent amplitude is
\beqr
\lan x(t) \ran & \approx & \bt_K \theta \frac{ Q}{[(1 - \xi^2(t) +  Q^2)^2 + 4 \xi^2(t)]^{3/4}}
\sin[\Phi(t) + 3\Theta(t) - \half q ]  \label{eq: pulse_nonlinq}
\eeqr
The echo amplitude at $t=2\tau$ is approximated by 
\beq
 \lan x(t=2\tau) \ran^{amp} \approx \bt_k \theta  \frac{Q}{(1 + Q^2 )^{3/2}} 
\label{eq: amp_nonlinq}
\eeq
This expression has the same form as Eq.~(4.10) in \cite{Stup_Kauf} evaluated at 
the time of the first echo. We expect however 
that the general form in Eq.~(\ref{eq: amp_nonlinq_gen}) will be more accurate for larger values of $q$. Finally we 
recover the completely linear theory by dropping the $Q^2$ term. In this case
$ \Theta(t) \approx {\rm Arctan}[\xi(t)]$  and we have
\beq
\lan x(t) \ran_{linear} = 
\bt_K \theta \frac{Q}{[(1 + \xi^2(t) ]^{3/2}} \sin (\Phi(t) + 3{\rm Arctan}[\xi(t)] - \half q )
\label{eq: amp_linear}
\eeq
Eq.~(\ref{eq: amp_linear}) is the same  as that obtained in \cite{Chao}, with the addition of the small correction to the
phase. The range of values in the quadrupole strength $q$ 
over which the linear theory is valid decreases as either the emittance or the dipole kick increases. 

In order to obtain the optimum quadrupole strength that maximizes the echo amplitude, 
we define a dimensionless parameter $\eta= \om' \eps_0 \tau = \tau/\tau_D$  in terms of which  $Q=q\eta$. Let $\sg_0  = \sqrt{\bt \eps_0}$ denote the rms beam size at a location with beta function
$\bt$. Then $\eta$ is the additional change in phase due to the nonlinearity of particles at the rms beam size
accumulated in the time between the two kicks. The optimum quadrupole  kick
$q_{opt}$ at which the echo amplitude reaches a maximum when $\eta \gg 1$ is given by
\beq
\lim_{\eta \gg 1} q_{opt} = \frac{1}{\sqrt{2}\eta} = \frac{1}{\sqrt{2}}\frac{1}{\om'\eps_0 \tau}   
\label{eq: qopt_NLq}
\eeq
Proceeding with the above form for $q_{opt}$, and substituting back into the simpler Eq.~(\ref{eq: amp_nonlinq}), 
the echo amplitude relative to the dipole kick  at the optimum quadrupole  strength
\beq
 \lim_{\eta \gg 1} A_{max} \equiv \lim_{\eta \gg 1} \frac{\lan x(2\tau) \ran^{max,amp}}{\bt_K\theta} = 
 \frac{2}{3\sqrt{3}} = 0.38   \label{eq: max_amp_1}
\eeq
The results for $q_{opt}$ and $A_{max}$ in this approximation of keeping only $T_1$ were first obtained in \cite{Stup_Kauf}. 
In this form, the maximum relative amplitude $A_{max}$ is a constant, independent of the initial emittance and dipole 
kick. We 
expect this to be true when the initial emittance is sufficiently large. 
We note that the value of $A_{max}$ observed with gold ions with their nominal emittances during the RHIC experiments
\cite{Fischer_2005}  was 0.35, close to this predicted value. 
Numerical evaluation of the complete amplitude function $A_{1,3}$ defined in Eq.~(\ref{eq: A13}) leads to a correction of 
about 10\% from that in Eq.~(\ref{eq: max_amp_1}). The simulations to be discussed in Section \ref{sec: simul} will
show that $A_{max}$  exceeds the above prediction for small emittances.

The above discussion has assumed that the rms angular betatron frequency spread is given by $\sg_{\om} =\om'\eps_0$. 
However, the beam  decoheres following the dipole kick and  the emittance grows from $\eps_0$ to 
$\eps_f= \eps_0[1 + \half(\bt_K \theta/\sg_0)^2]$ at times $t \gg \tau_D$ \cite{Ed_Sy, Sen_2017}. 
At these times, we assume that the increased rms frequency spread can be approximated by $\sg_{\om} \approx \om' \eps_f$. 
In the next section, we will calculate this rms frequency spread exactly and show that this approximation is valid in the limit of small amplitude dipole kicks $\bt_K \theta \ll \sg_0$. 
Incorporating this increased emittance and frequency spread had turned out to be essential in comparing
theory with the experimental measurements at RHIC \cite{Sen_2017}.
We can include these effects into the above equations by the approximate modifications
\beqr
\xi(t) &  \approx & (t - 2\tau)\om' \eps_0[1 + \half (\frac{\bt_K \theta}{\sg_0})^2 ]  \label{eq: mod_xi}\\
Q & \approx & q \tau \om' \eps_0 [1 + \half (\frac{\bt_K \theta}{\sg_0})^2 ]
\label{eq: mod_Q}
\eeqr
These changes lead to a theory which is nonlinear in the dipole kick but this is an incomplete 
dependence. A more complete nonlinear theory will be discussed in the next section.

The plots in Figure \ref{fig: theory1} show the echo amplitude dependence on the quadrupole  kick, as predicted by 
Eq.~(\ref{eq: amp_nonlinq_gen}) with and without the modifications introduced in Eqs. (\ref{eq: mod_xi})  and (\ref{eq: mod_Q}).  
For a very small initial emittance (left plot), the black curve shows that the relative echo amplitude without emittance growth is 
independent of the dipole kick and increases monotonically with the quadrupole kick; the relative amplitude reaches nearly 
0.5 at $q=1$. The blue and red curves for
dipole kicks of 1mm and 3mm respectively include the increased frequency spread which changes the profiles significantly. 
In both cases, $A_{max}$ is close to 0.38, while $q_{opt}$ shifts to lower values. The right plot in Fig.~\ref{fig: theory1} shows 
results with a larger initial emittance chosen close to measured values in the RHIC experiments \cite{Fischer_2005}. 
In this case, 
even without including the increased emittance from the dipole kick, $A_{max}$ does not exceed 0.38. Including
the increased frequency spread shifts $q_{opt}$ to lower values, as expected since $q_{opt} \propto 1/\sg_{\om}$. 
The two plots combined also show that  $q_{opt}$ decreases with increasing emittance. 
\bfig
\centering
\includegraphics[scale=0.55]{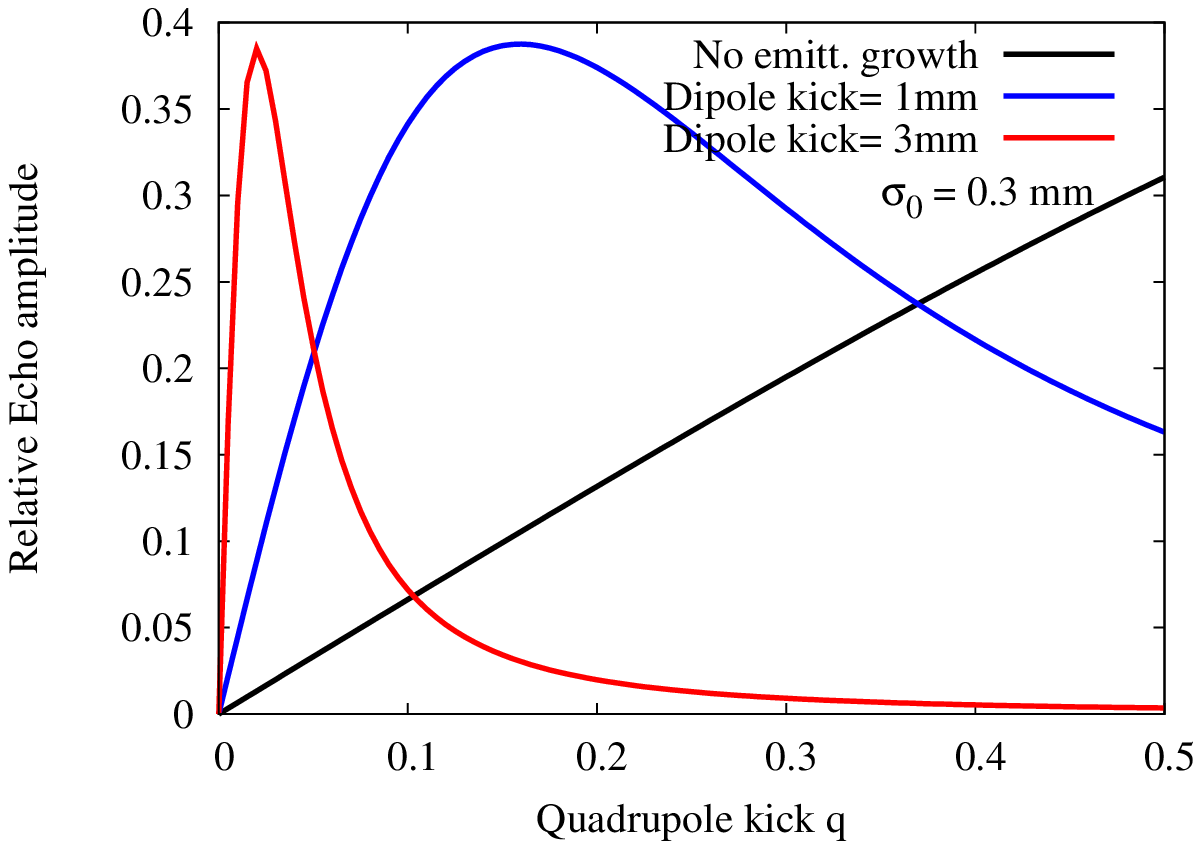}
\includegraphics[scale=0.55]{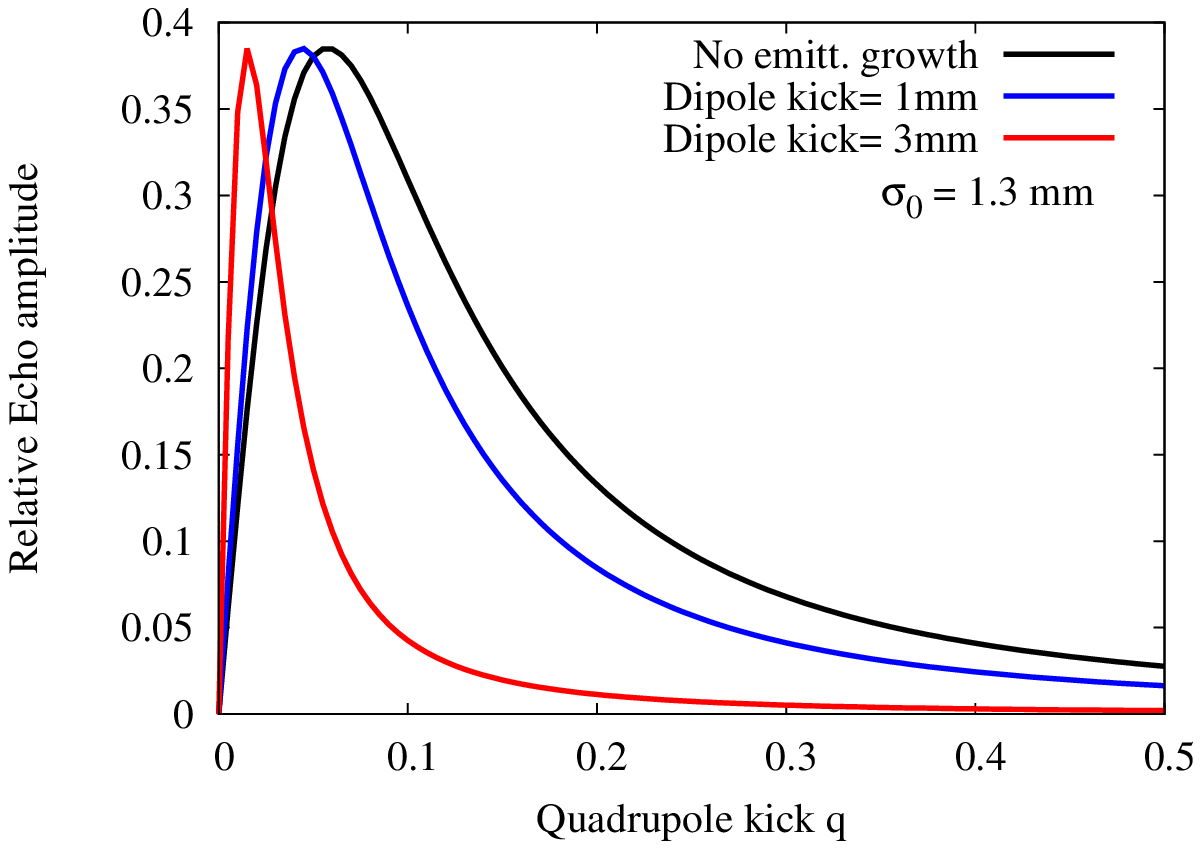}
\caption{The echo amplitude relative to the dipole kick as a function of quadrupole  kick predicted by Eq.~(\ref{eq: amp_nonlinq_gen}). The two 
plots are for different initial emittances. In each plot, the black curve shows the prediction without including the emittance 
growth
from a dipole kick, the blue and red curves include emittance growth from dipole kicks of 1mm and 3mm respectively. }
\label{fig: theory1}
\efig

%\clearpage

\setcounter{equation}{0}
\section{Nonlinear theory of dipole and quadrupole kicks}  \label{sec: NL_dip}

There are a few drawbacks to the theory developed in the previous section. The first is that it has an incomplete
dependence on the dipole kick strength; the emittance growth had to be introduced as a correction. The value of $A_{max}$
is limited to 0.38, in disagreement with simulation results.  It also does not predict the existence of multiple echoes at times beyond the first
one at 2$\tau$. However the experiments at RHIC (cf. Fig.~5 in \cite{Fischer_2005}) showed echoes at 4$\tau$ and 6$\tau$.
These multiple echoes are also seen in simulations, see. Fig.~\ref{fig: mechoes_dip1-6_em3} in Section \ref{sec: simul}. 
Our aim is to develop a 
theory, labeled DQT, that is nonlinear in both dipole and quadrupole  strengths which will remove these drawbacks.

In this section, we will make an approximation for the change in the distribution function that includes the large time
dependent change in the angle $\phi$ but neglects the smaller impulsive changes to the action and angle. This results in 
expressions which are
approximate but contain the essential physics. The more complete theory which results in more complicated expressions
 is developed in Appendix A.

Using the notation of Section \ref{sec: NL_quad}, the complete distribution function (DF) without the first order Taylor 
expansion at time $t$ after the dipole kick is
\beqr
\psi_2(J,\phi,  t) = \psi_0(J + \bt_K\theta \sqrt{2 J/\bt}\sin \phi_{-t}+ (1/2)\bt_K\theta^2) , \;\;\;
\phi_{- t} \equiv \phi - \om(J) t    \nonumber \\
\mbox{}  \label{eq: psi3_complete} 
\eeqr
This DF can be used to calculate the increased emittance and the tune spread after the dipole kick. The time dependent rms emittance was calculated in \cite{Sen_2017}. 
The action dependent frequency spread is $\Dl \om = \om' J$ from which the rms frequency spread $\sg_{\om}$ can be found from
\beq
\sg_{\om} = \sqrt{\lan (\Dl \om)^2 \ran - (\lan \Dl \om \ran)^2 }, \;\;\; \lan \Dl \om\ran = \om' \int dJ \; J\int d\phi \psi_2(J, \phi, t) 
\eeq
Using the form of $\psi_2$ in Eq. (\ref{eq: psi3_complete}), we find that the exact rms frequency spread after the dipole kick 
is
\beq
\sg_{\om} = \om' \eps_0 \sqrt{1 + \frac{\bt_K \theta^2}{\eps_0} }
\eeq
In the limit of small amplitude dipole kicks, this reduces to the approimate form assumed in Eq.(\ref{eq: mod_xi}) and 
Eq. (\ref{eq: mod_Q}). The increase in the frequency
spread leads to a smaller decoherence time after the dipole kick, as will also be seen in the simulations. 

We follow the same transformations as in Section  \ref{sec: NL_quad} to calculate the centroid motion following the quadrupole kick at time $t=\tau$.
 The dominant contribution to the change in the DF after the quadrupole kick at time $t > \tau$ is the transformation 
due to the angle evolution 
$\phi_{-\tau}  \rarw  \phi_{-\Dl\phi} -  \tau \om + Q (J/\eps_0) \sin 2\phi_{-\Dl \phi}$ because these 
grow with time as $(t-\tau)$ and the delay $\tau$ ($Q$ depends on $\tau$). 
The theory in Appendix A includes the smaller transformations due to the impulsive kicks. 
Here, in the approximation of keeping only this dominant term, 
 the DF as a function of the scaled action variable $z = J/\eps_0$ at a time after the quadrupole kick  $t > \tau$ is 
\beqr
\psi_5(z, \phi,t) & = & \psi_0(z\eps_0 + \bt_K\theta\sqrt{\frac{2 \eps_0 z }{\bt}}
\sin(\phi_{- \Dl\phi}  - \tau \om - q\cos^2\phi_{-\Dl\phi} + Qz\sin 2\phi_{-\Dl\phi}) + 
\half \bt_K\theta^2 )  \nonumber \\
\mbox{} 
\eeqr
  We have for the dipole moment
\beqr
\lan x(t) \ran & = & \frac{\sqrt{2\bt \eps_0}}{2\pi} \exp[-\frac{\bt_K\theta^2}{2 \eps_0}]\int dz \; \sqrt{z} \exp[- z] 
T_{\phi}(z) \\
T_{\phi}(z)  & \simeq &  {\rm Re} \left\{  \int d\phi e^{i\phi} \exp\left[ - a_{\theta}\sqrt{2z} \sin( \phi_{- \Dl\phi} - \tau \om - 
\half q + Q z \sin 2\phi_{-\Dl \phi} ) \right] \right\}
\eeqr
where we introduced the dimensionless dipole kick parameter in units of the rms beam size
\beq
a_{\theta} = \fr{\bt_K \theta}{\sqrt{\bt \eps_0}}
\eeq

One way of doing the $\phi$ integration is to use the generating functions for the modified Bessel function $I_n(z)$ and
for the Bessel function $J_n(z)$ \cite{Abram_Steg}, i.e.
\[  e^{-z\sin\theta} = \sum_{n=-\infty}^{\infty}i^n I_n(z) e^{i n \theta}, \;\;\;  e^{i z \sin\theta} = \sum_{l=-\infty}^{\infty} J_l(z) e^{i l \theta} \]
Then the term $T_{\phi}(z)$ transforms to
\beqr
T_{\phi}(z) & = &  {\rm Re} \left\{  \sum_{k=-\infty}^{\infty} \sum_{l=-\infty}^{\infty} i^{k}  I_{k}(a_{\theta}\sqrt{2 z}) J_l(k Qz)\exp[i(- k(\Dl\phi + \tau\om + \half q) - 2l\Dl\phi) ]  \right. \nonumber \\
& & \left. \times  \int d\phi  \exp\left[ i ( [1 + k  + 2l ] \phi)  \right]  \right\} \nonumber \\
& =  &  2\pi {\rm Re} \left\{  \sum_{l} i^{-(2l+1)}  I_{-(2l+1)}(a_{\theta}\sqrt{2 z}) J_{l}(-(2l+1) Qz) \right. \nonumber \\
&  &  \left. \times \exp\left[ i(\om (t + 2l \tau) + \half (2l+1) q)\right] \right\}
\eeqr
Since the sum extends over positive and negative values of $l$, we can replace $l = -n$ and write
\beq
 \om(t - 2 n \tau) \equiv \Phi_n + \xi_n z , \;\;\; \Phi_n= \om_{\bt}(t - 2 n \tau), \;\;\;  \xi_n =  \om'\eps_0(t - 2n \tau)
\eeq
We have therefore for the time dependent echo pulse
\beqr
\lan x(t) \ran  &  =  &  \sqrt{2\bt \eps_0} e^{-(\bt/2\bt_K)a_{\theta}^2}
  {\rm Im} \left\{  \sum_{n=-\infty}^{\infty}   \exp[i( \Phi_n - \half (2n - 1)q )]  \right.   \nonumber \\
&  & \left.  \times \int dz \;  \sqrt{z} \exp[- z (1 - i\xi_n ) ] I_{2n - 1}(a_{\theta}\sqrt{2 z}) J_{n}([2n - 1] Qz)  \right\} 
\label{eq: echo_t_NLdip_1}
\eeqr
where we used  ${\rm Re}[ - i f(z)] = {\rm Im}[f(z)]$ for a complex function $f(z)$. 
This echo pulse will be large when the dominant phase factors  $\Phi_n = 0 = \xi_n$, i.e at times $t = 2 n\tau$.
This form therefore predicts  echoes at times close to multiples of 2$\tau$. The presence of the small $q$ dependent phase 
factor i.e. $(2n-1)q/2$  will shift the maximum of the
echo away from $2n\tau$, the shift increasing with $q$ and the order $n$ of the echo. 
The dipole moment of the first echo ($n=1$), under the approximations made in this section, is
\beqr
\lan x(t=2\tau) \ran  &  =  &  \sqrt{2\bt \eps_0} e^{-(\bt/2\bt_K)a_{\theta}^2}
{\rm Im} \left\{ e^{i(\Phi_1 - q/2)} \right. \nonumber \\ 
&  &  \left. \times \int dz \;  \sqrt{z} \exp[- z\{1 - i \xi_1 \} ] I_1(a_{\theta}\sqrt{2 z}) J_{1}( Qz) \right\}
\label{eq: 2tau_NLDip_1}
\eeqr
This form can be compared with the term $T_1$ in Section \ref{sec: NL_quad} in the linear dipole 
approximation, which  was (before the integration over $z$)
\beq
\lan x(t=2\tau) \ran_{QT}  = \bt_K\theta {\rm Im}\left\{ e^{i[\Phi_1- q/2]}\int dz \; z \exp[-z\{1 - i \xi_1 \}] J_1(Qz)  \right\}
\label{eq: NL_quad_T1}
\eeq
If in Eq.~(\ref{eq: 2tau_NLDip_1}) we replace $I_1(a_{\theta}\sqrt{2 z})$ by its first order approximation
 $\half a_{\theta}\sqrt{2 z}$ and $e^{-(\bt/2\bt_K)a_{f}^2}$ by 1, then it reduces to  Equation \ref{eq: NL_quad_T1}. 
In Section \ref{sec: NL_quad}, we  included the emittance growth due to the dipole kick  in a post hoc fashion
by changing $\eps_0$ to $\eps_f$ in parameters such as $\xi, Q$ etc.
In this section, the use of the complete distribution function to all orders in the dipole kick, e.g. $\psi_3$ in 
Eq.~(\ref{eq: psi3_complete}), naturally accounts for the emittance growth as is seen by calculating the second moments
\cite{Sen_2017}. Hence we use the original definitions of the parameters $\xi, Q$ in evaluating Eq.~(\ref{eq: 2tau_NLDip_1}).
This equation shows that the maximum relative echo amplitude depends on the relative dipole kick $a_{\theta}$ through 
$\exp[-(\bt/(2\bt_K))a_{\theta}^2] I_1(\sqrt{2}a_{\theta} \sqrt{z})$ and 
on the quadrupole strength $q$, the emittance $\eps_0$, the lattice nonlinearity $\om'$, and the delay $\tau$ through 
$J_1(q\om'\tau \eps_0 z)$.

The amplitude of the echo at 4$\tau$ corresponds to the term with $n=2$ in Eq.~(\ref{eq: echo_t_NLdip_1}). Hence
\beqr
\lan x(t=4\tau) \ran &   =  &  \sqrt{2\bt \eps_0} e^{-(\bt/2\bt_K)a_{\theta}^2} \nonumber \\
& & \times {\rm Im} \left\{ e^{i(\Phi_2 - 3q/2)}\int dz \;  \sqrt{z} \exp[-  z \{1 - i\xi_2  \}] I_3(a_{\theta}\sqrt{2 z}) 
J_{2}( Qz) \right\} \nonumber \\
\label{eq: 2tau_NLDip_2}
\eeqr
Note that since the lowest order term in $I_3(a_{\theta}\sqrt{2 z})$ is $(a_{\theta}\sqrt{z})^3 $, there is no
echo at 4$\tau$ in the linearized dipole kick approximation. 

The integrals in Eq.~(\ref{eq: 2tau_NLDip_1}) and Eq.~(\ref{eq: 2tau_NLDip_2}) do not appear to be analytically tractable nor do they
appear to be listed in the extensive tables of integrals in \cite{Grad_Ryz}.  However they can be evaluated 
numerically. As a consequence however, the optimum quadrupole strengths to maximize the echo amplitudes must be found
numerically, unlike the case with the theory developed in Section \ref{sec: NL_quad}.
Detailed comparisons of the predictions from QT and DQT theories are discussed in 
the next  section on simulations.

We briefly illustrate how the nonlinear nature of the dipole kicks changes the echo response. Fig.~\ref{fig: echo1_echo2} 
shows the impact of increasing dipole kicks on the amplitudes of the first and second echoes, based on the above theory. In 
general we find that increasing the dipole kick
lowers the optimum quadrupole kick $q_{opt}$ and increases the relative amplitude slightly, as also seen in Section 
\ref{sec: NL_quad}. On the other hand for the second echo, larger dipole kicks also decrease the corresponding
$q_{opt}$ but significantly increase its amplitude. 
\bfig
\centering
\includegraphics[scale=0.55]{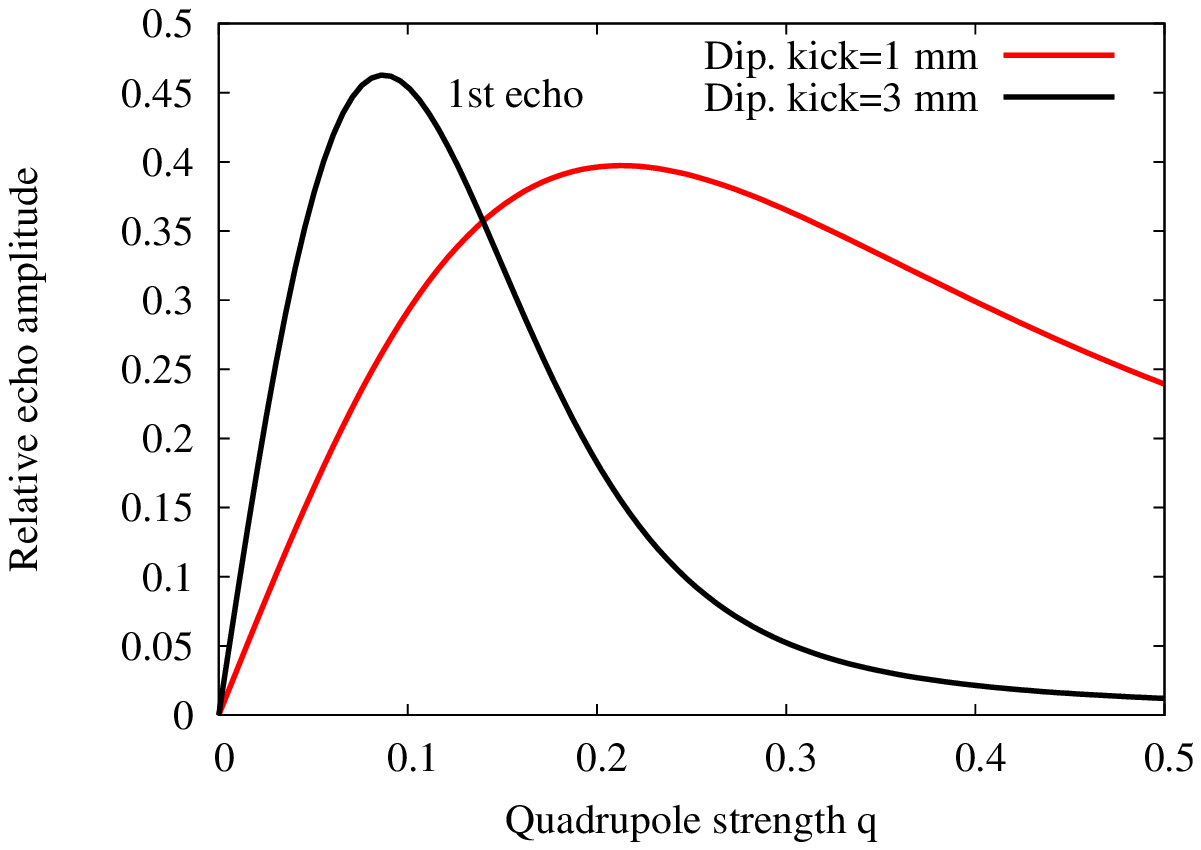}
\includegraphics[scale=0.55]{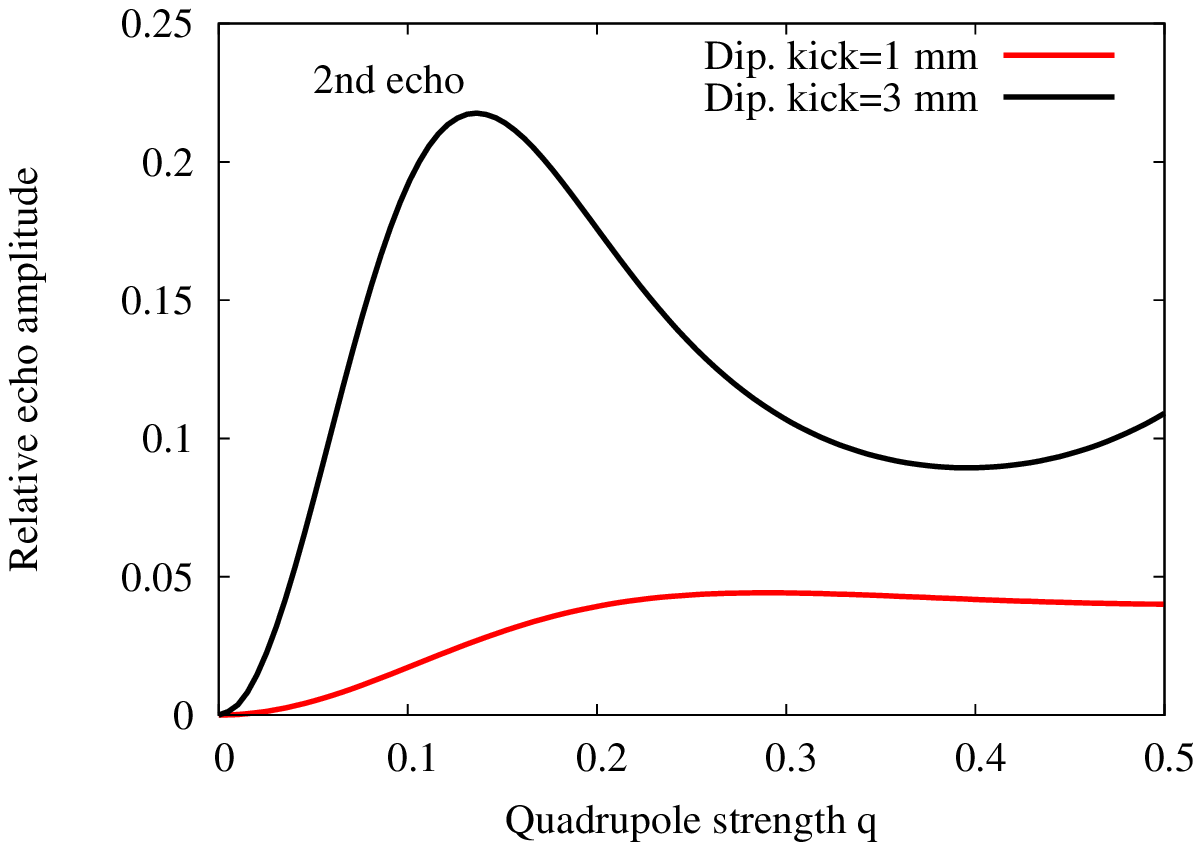}
\caption{Left: Relative amplitude of the first echo vs quadrupole strength for two dipole kicks. Right: Relative amplitude of the
second echo for the same two dipole kicks. The initial emittance is the same in both plots. }
\label{fig: echo1_echo2}
\efig
The left plot in this figure shows the first echo's amplitude $A^{(1)}$ as a function of the quadrupole kick $q$ for two 
dipole kicks at a constant beam size of 1mm. As the dipole kick increases from 1 mm to 3 mm, $q_{opt}$ decreases 
while the echo amplitude at $q_{opt}$ increases slightly. 
The right plot shows the response of the second echo as a function of $q$. At a 1 mm dipole kick, the second echo's amplitude  
$A^{(2)}$ has a relatively flat response to the quadrupole kick after an initial linear increase. At a 1mm kick, 
$A_{max}^{(2)} \sim 0.1 A_{max}^{(1)}$ while at a 3mm dipole kick $A_{max}^{(2)} \sim 0.5 A_{max}^{(1)}$.
Increasing the dipole kick shows that $A_{max}^{(1)}(\bt_K \theta = 3 \; {\rm mm}) \sim 1.15 A_{max}^{(1)}(\bt_K \theta = 1 \; {\rm mm})$ while $A_{max}^{(2)}(\bt_K \theta = 3 \; {\rm mm}) \sim 5 A_{max}^{(2)}(\bt_K \theta = 1 \; {\rm mm})$,
showing that the second echo is much more sensitive to the dipole kick.
Summarizing,  we have shown that the nonlinear dipole and quadrupole theory (DQT) removes the drawbacks
of the nonlinear quadrupole theory (QT) mentioned earlier.

%\clearpage

\setcounter{equation}{0}
\section{Simulations of echo amplitudes} \label{sec: simul}

In this section, we discuss the results of 1D echo simulations using a simple particle tracking code. The code models linear motion in an accelerator ring and nonlinear motion due to octupoles placed around the ring. A single turn dipole kick acts on the particle distribution at a chosen moment and is followed by a single turn quadrupole kick at a later time after the distribution has decohered. The beam distribution is then followed at a separate observation point for a virtual beam position monitor (BPM) and the first 
moment is recorded until the first few echoes have developed and then disappeared. The main beam parameters in the simulations are shown in Table \ref{table: parameters}. We do not specify the beam energy here, but  note
that the emittances chosen are in a range around the nominal un-normalized emittance observed 
during 100 GeV operation with proton beams at RHIC \cite{Fischer_2005}. 
The octupole strengths were chosen to ensure a large enough nonlinear tune spread that results in 
decoherence times of the order of
a few hundred turns but small enough that no particles were lost at the largest dipole kick used. Typically, the dipole kick was applied after 200 turns and the quadrupole kick at turn 1600. This delay time of 1400 turns is large enough so that 
the beam distribution had decohered completely (in most cases, but see the discussion below) at the time of the quadrupole kick. 
A Gaussian beam distribution in transverse  $(x,p)$ space  with three seeds for each echo simulation was used and averaged to 
obtain the echo amplitude. The simulations were
done for different initial emittances, dipole kicks and quadrupole kicks while keeping the detuning and delay parameters constant. 
\begin{table}
\bec
\btable{ccc} \hline
Parameter & Symbol & Value  \\  \hline
Number of particles & $N_{part}$ & 20000 \\
Total simulation turns &  - & 4000 - 10,000\\
Tune & $\nu_{\bt}$ & 0.245 \\
Beta function at BPM, dipole, quadrupole [m] & $\bt, \bt_K, \bt_Q$ & 10, 10, 10 \\
Dipole kick  range [mrad] & $\theta$ &  0.1 - 1.0 \\
Quadrupole kick range & $q$ &  0.01 - 0.5 \\
Delay time [turns] & $N_{\tau}$ & 1400 \\
Tune slope [1/m] & $\nu'$ & -3009 \\
\hline
\etable
\eec
\caption{Table of parameters}
\label{table: parameters}
\end{table}

First, we make some general observations. The emittance growth following the dipole kick was compared with the 
prediction,   $\eps_f = \eps_0[1 + \half(\frac{\bt_K \theta}{\sg_0})^2] $
and found to be within 5\% of this value. Also as expected,  there was no further emittance growth 
following the quadrupole kick. The decoherence  time, calculated as the e-folding time for the centroid decay following the
dipole kick, depends both on  the initial emittance and on the dipole kick. 
 We also observe that for small emittances and small 
dipole kicks where the decoherence time is longer than 1400 turns, the quadrupole kick was applied before the beam had
completely decohered. Echoes are still observed, albeit of relatively small amplitude. These echoes have long durations that are proportional 
to the decoherence time; as predicted by the linear theory \cite{Chao}. 

Figure \ref{fig: centroid_em1_dip3} shows an example of the change in the echo 
pulse shape with increasing values of $q$, at constant emittance and constant dipole kick. 
We observe that as low values of $q $, the echo pulse is symmetric and increases in amplitude with $q$, but with further increase
becomes asymmetric, widens, starts earlier than $2\tau$, then  splits into two pulses of smaller  amplitudes before vanishing 
altogether. The plots in Figure 
\ref{fig: centroid_em1_q0.1} show  the echo pulse with increasing dipole kicks, at constant 
initial emittance and constant quadrupole kick. The plots in Fig.~\ref{fig: centroid_em1_q0.1} also illustrate that the 
decoherence time decreases as the dipole kick increases. The first plot in this figure shows that an echo pulse is still formed, 
even though the centroid has not completely decohered at the time of the quadrupole kick. 
\bfig
\centering
\includegraphics[scale=0.35]{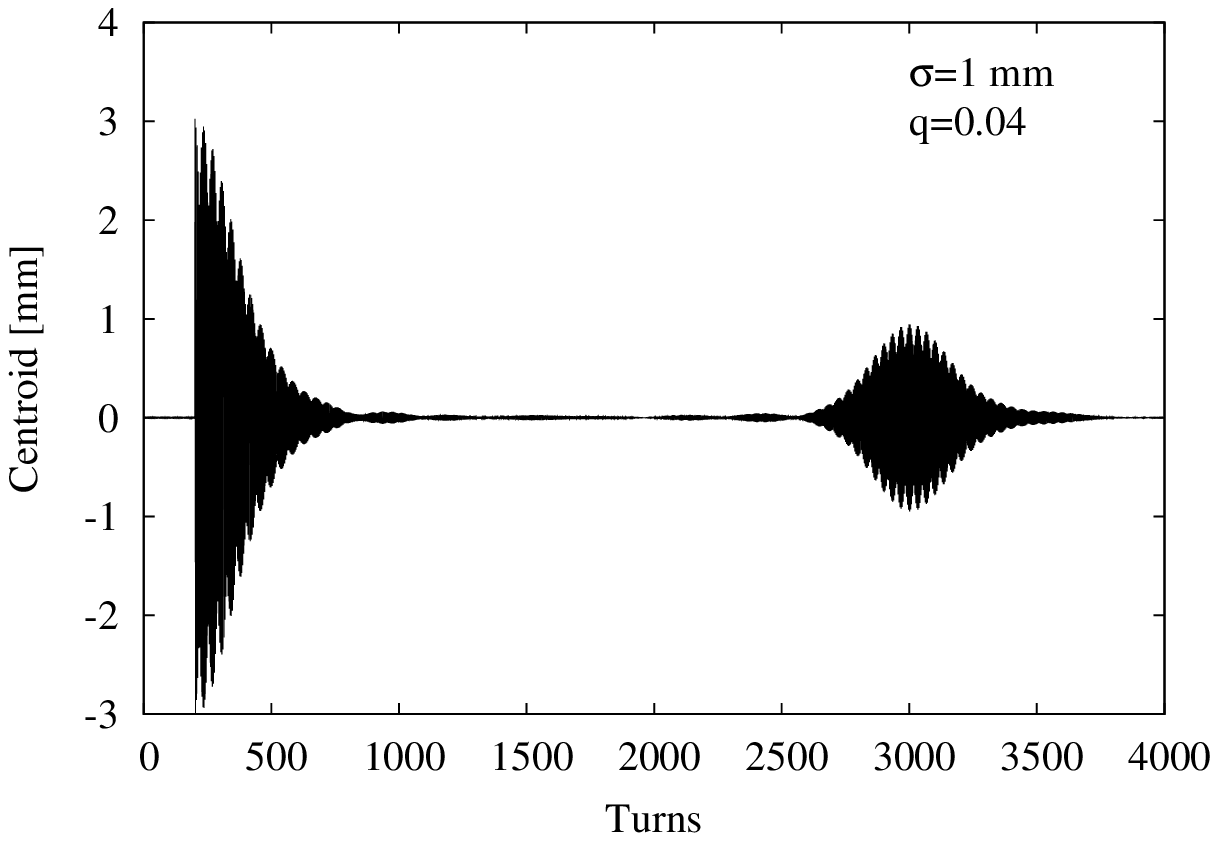}
\includegraphics[scale=0.35]{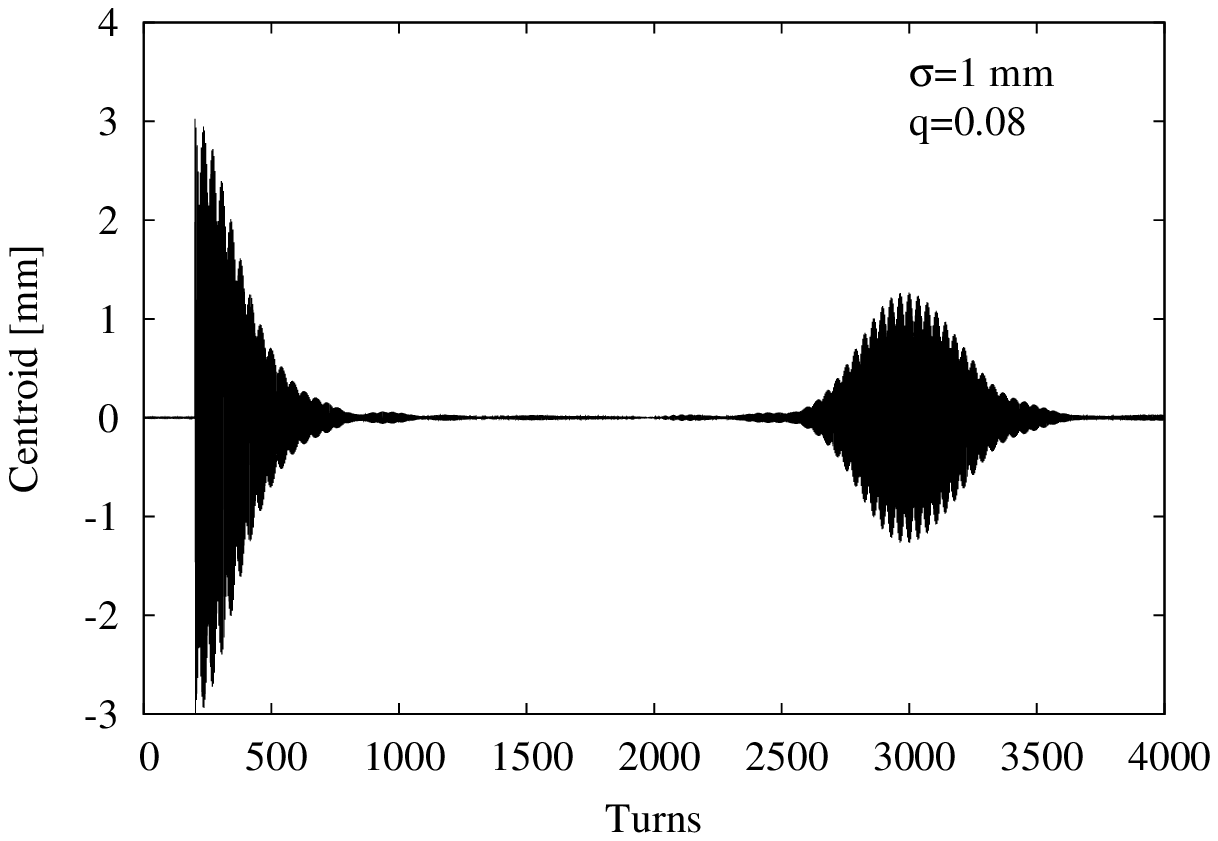}
\includegraphics[scale=0.35]{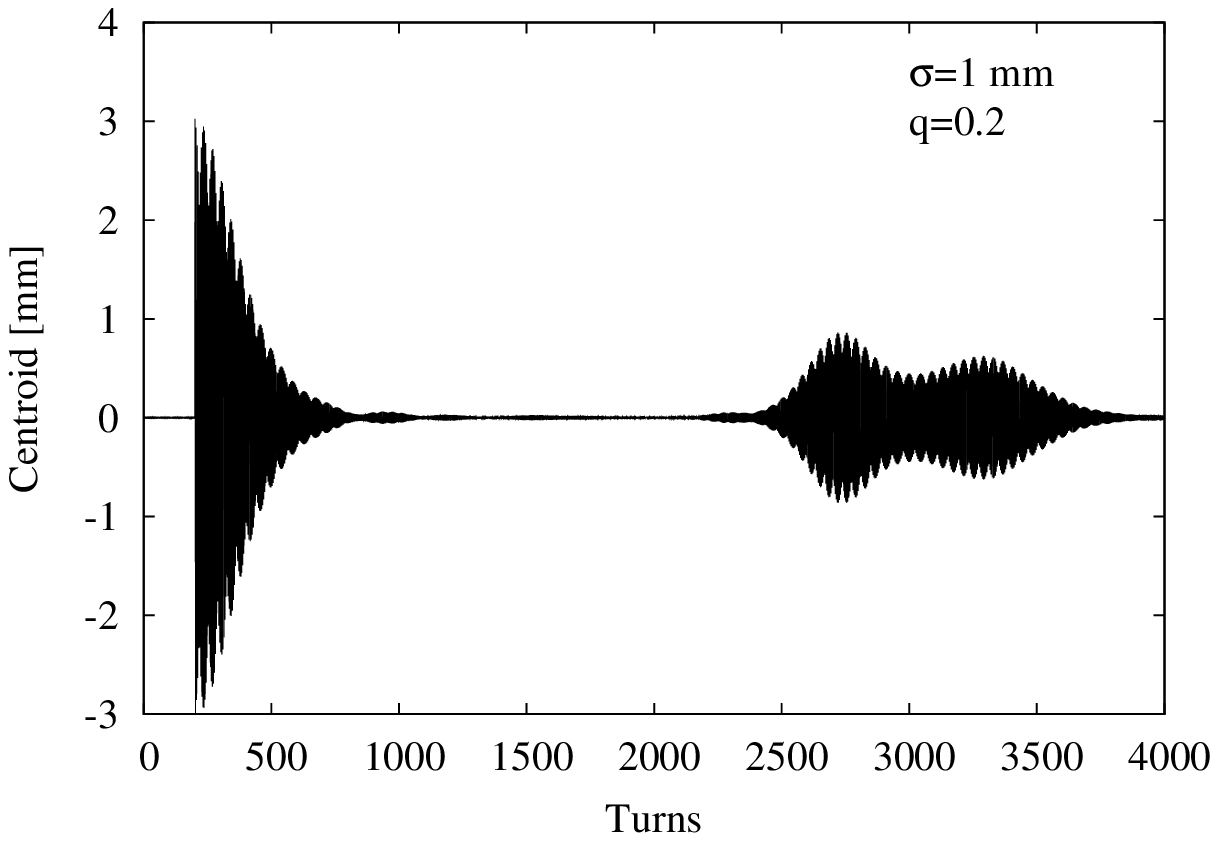}
\caption{Time evolution of the centroid after the dipole kick at turn 200, with different strength quadrupole kicks applied at
turn 1600.  The echo pulse is centered around turn 3000.  Both the initial emittance (corresponding to $\sg_0=1$ mm at the BPM) 
and dipole kick = 3 mm were kept constant.  Quadrupole kicks 
increase from left to right in the three plots. The relative amplitude of the echo has a maximum at $q=0.08$ (center plot) at 
the chosen emittance.  }
\label{fig: centroid_em1_dip3}
\centering
\includegraphics[scale=0.35]{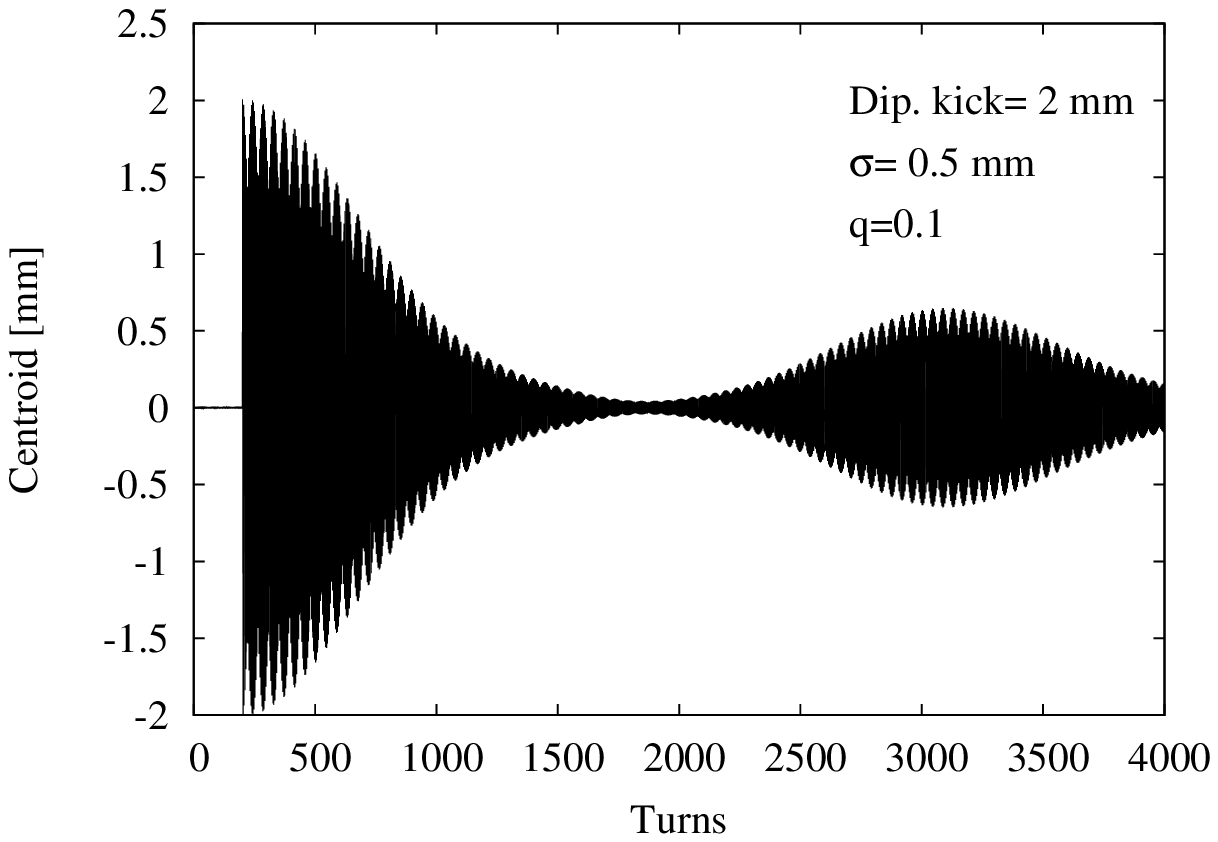}
\includegraphics[scale=0.35]{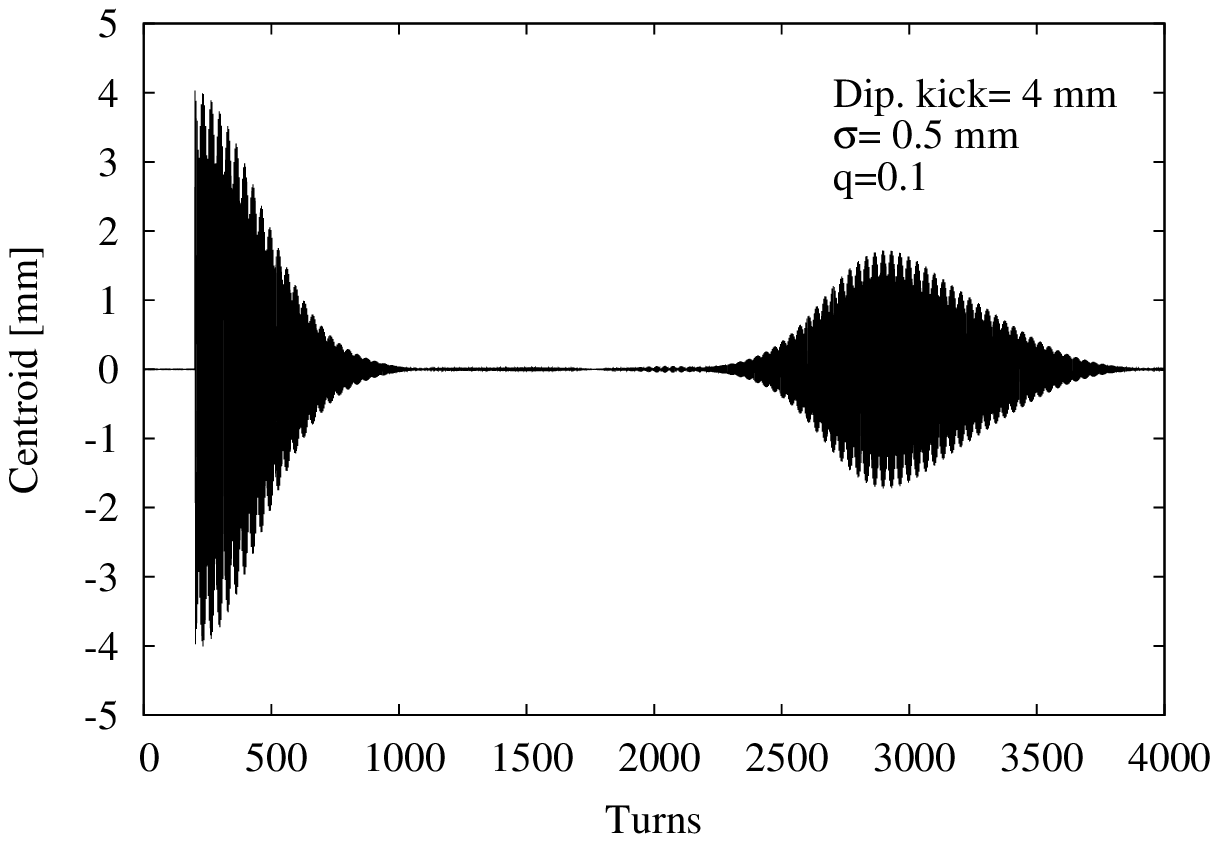}
\includegraphics[scale=0.35]{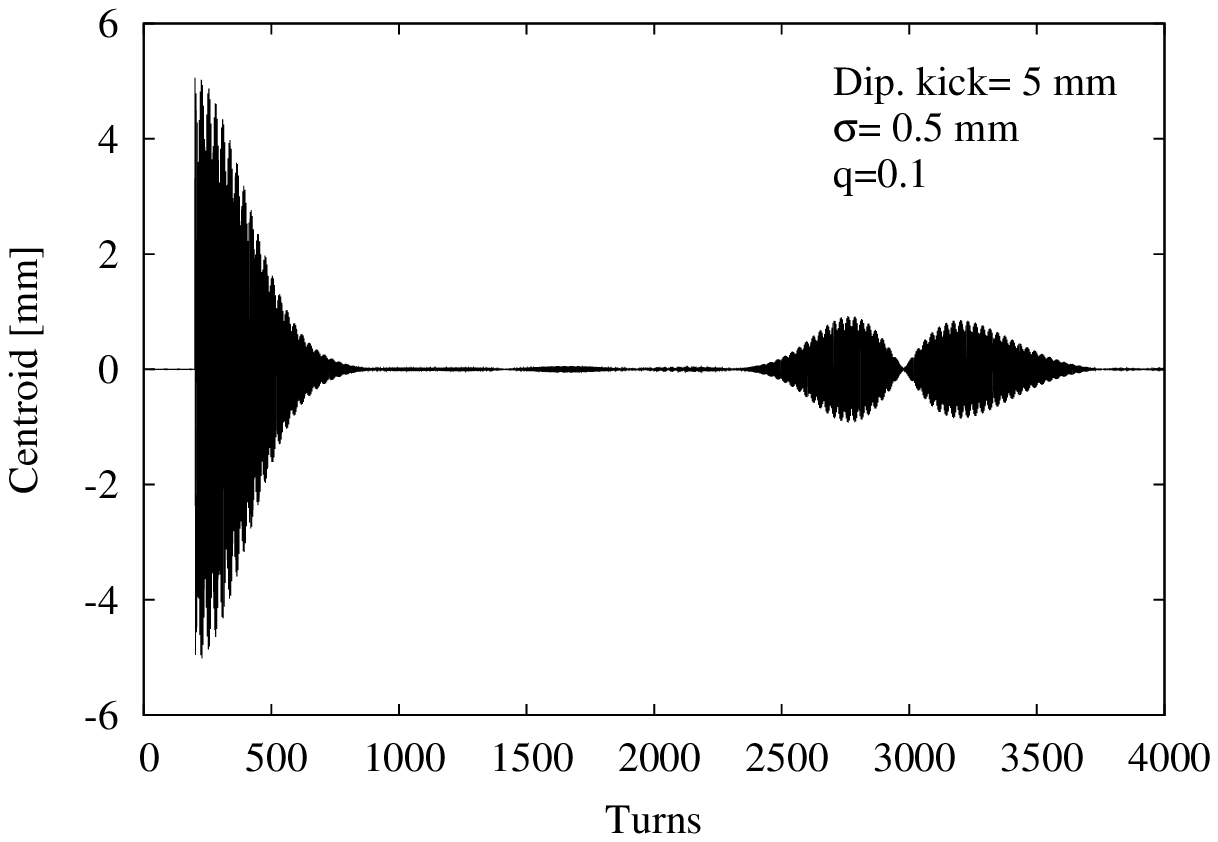}
\caption{Centroid evolution with different dipole kicks,  constant constant emittance ($\sg=0.5$mm), and constant 
quadrupole kick $q=0.1$. The dipole kicks increase from left to right. The decoherence time decreases with increasing
dipole kick. }
\label{fig: centroid_em1_q0.1}
\efig

Our goal is to maximize the echo signal by proper choices of parameters.
Fig.~\ref{fig: amp_q_em} shows theory and simulations of the echo amplitude as a function
of the quadrupole  strength for different values of the beam emittance and the initial dipole kick. Here we will consider the 
simpler version of the nonlinear dipole quadrupole theory (DQT) developed in Section \ref{sec: NL_dip}. 
The error bars represent the rms variation over the  seeds for the initial beam distribution and are quite small in every case. 
First we make general comparisons between the two theories with the simulation results. The nonlinear dipole and quadrupole
 theory (DQT) predicts larger amplitudes than the nonlinear quadrupole theory (QT) and is usually in better agreement with the simulations. The QT predicts
$A_{max}\le 0.4$, but the DQT and the simulations show larger values of $A_{max}$, especially at smaller
emittances. The QT predicts that the optimum $q_{opt}$ is determined by $\eta$, the
ratio of the delay to the decoherence time (see Eq.~(\ref{eq: qopt_NLq})). However the DQT 
predicts sligthly larger values of $q_{opt}$ than the QT, and that the echo amplitude  decreases more slowly for $q > q_{opt}$. 
All of these predictions from DQT are in better agreement
with the simulations. The differences between the theories  diminish with increasing emittance. At the larger emittances
studied,  $A_{max}$ in the simulations does not exceed 0.38, in agreement with the prediction of QT.

Now we turn to specific comparisons of the results shown in Fig.~\ref{fig: amp_q_em} where the initial emittance increases from top to bottom and the dipole kick increases from left to right. The top left plot in Fig.~\ref{fig: amp_q_em}, shows that the simulation points are at larger amplitude than the theories. In this case
the decoherence time is very long, so there is a contribution from the  initial 
dipole kick to the centroid amplitude at the time of the echo. Consequently, the simulated echo amplitude appears to be non-zero 
at zero quadrupole kick. The top right plot for the larger dipole kick (3mm) shows that the peak echo amplitude from simulation lies in
between the peak amplitudes from the QT and the DQT. Theoretical values of the optimum $q_{opt}$ are close to the
simulation value. However, the DQT shows a spurious oscillation for $q > 0.25$ at this low emittance. This occurs because of
the oscillatory integrand and the simple numerical integration algorithm used which does not converge rapidly enough in this 
parameter range where both $a_{\theta}$ is large and $q \gg q_{opt}$. There are straightforward algorithms
to improve the convergence with Bessel function integrands, see for example \cite{Lucas} . Such an algorithm can be
implemented if required. 
The plots in the second and third row show that for larger emittances, both theories (especially the DQT) agree reasonably well with simulations for
$q < q_{opt}$ but fall off faster with increasing $q$  for $q > q_{opt}$ compared to the simulations. 
 These differences may not be practically relevant, since we will use quadrupole kicks as close as possible to the optimum in experiments. In addition, the discrepancies for $a_{\theta} \ge 6$ may practically not matter, since it
is unlikely that the beam will be kicked to amplitudes larger than $6\sg$, especially in hadron superconducting
machines or in machines with collimator jaws placed close to this amplitude. 
\bfig
\centering
\includegraphics[scale=0.55]{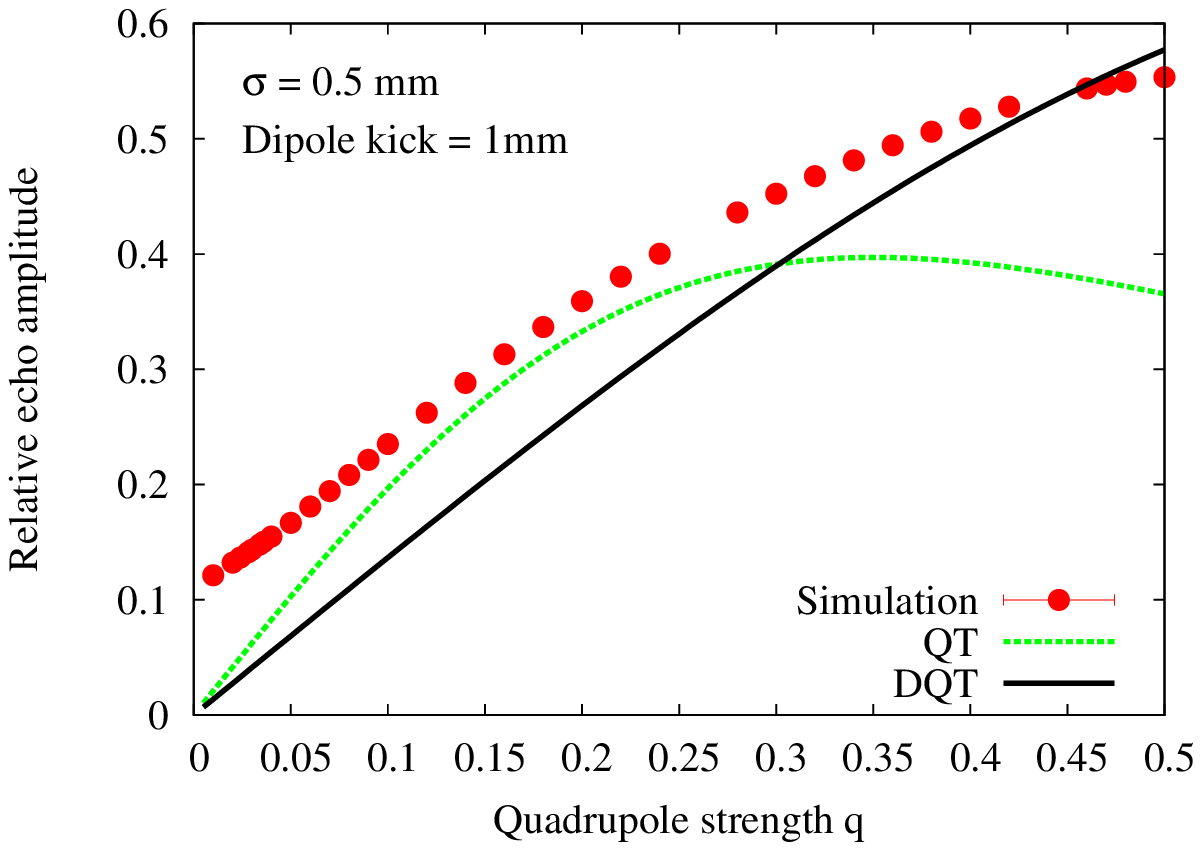}
\includegraphics[scale=0.55]{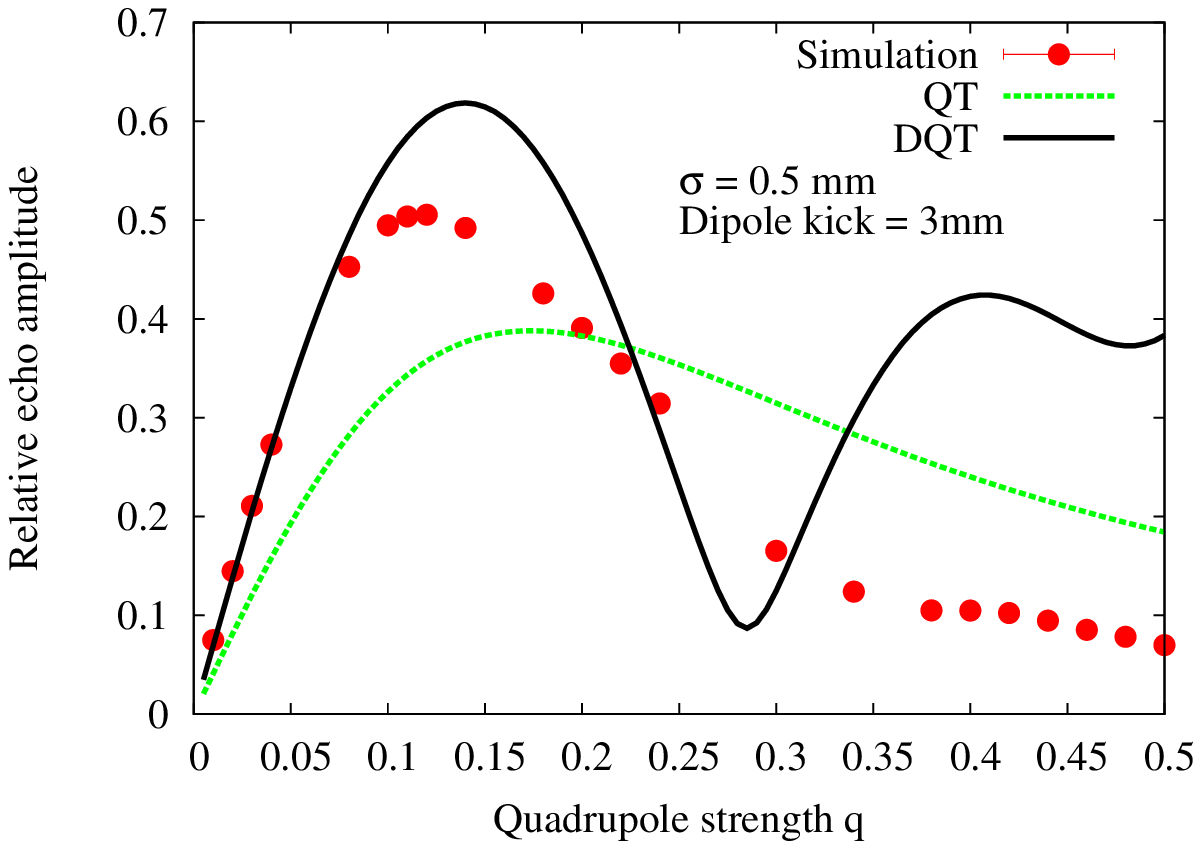}
\includegraphics[scale=0.55]{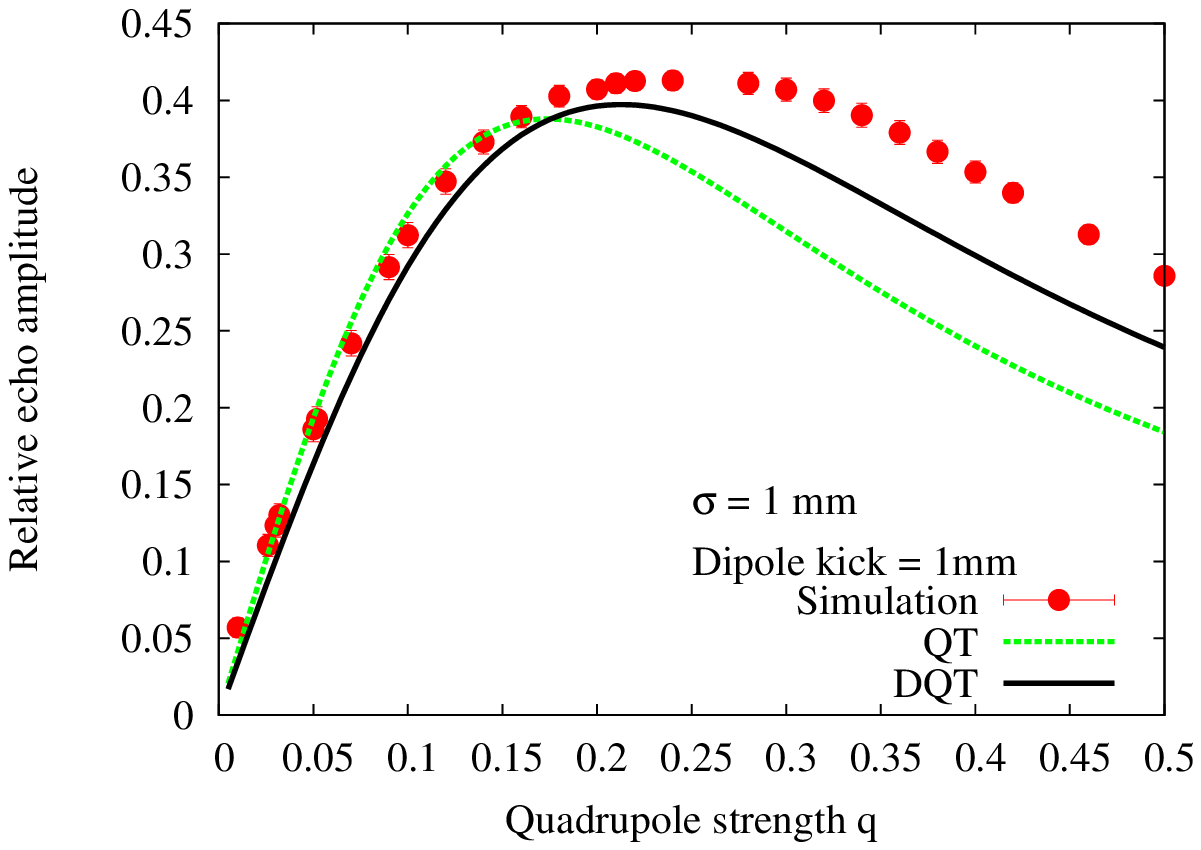}
\includegraphics[scale=0.55]{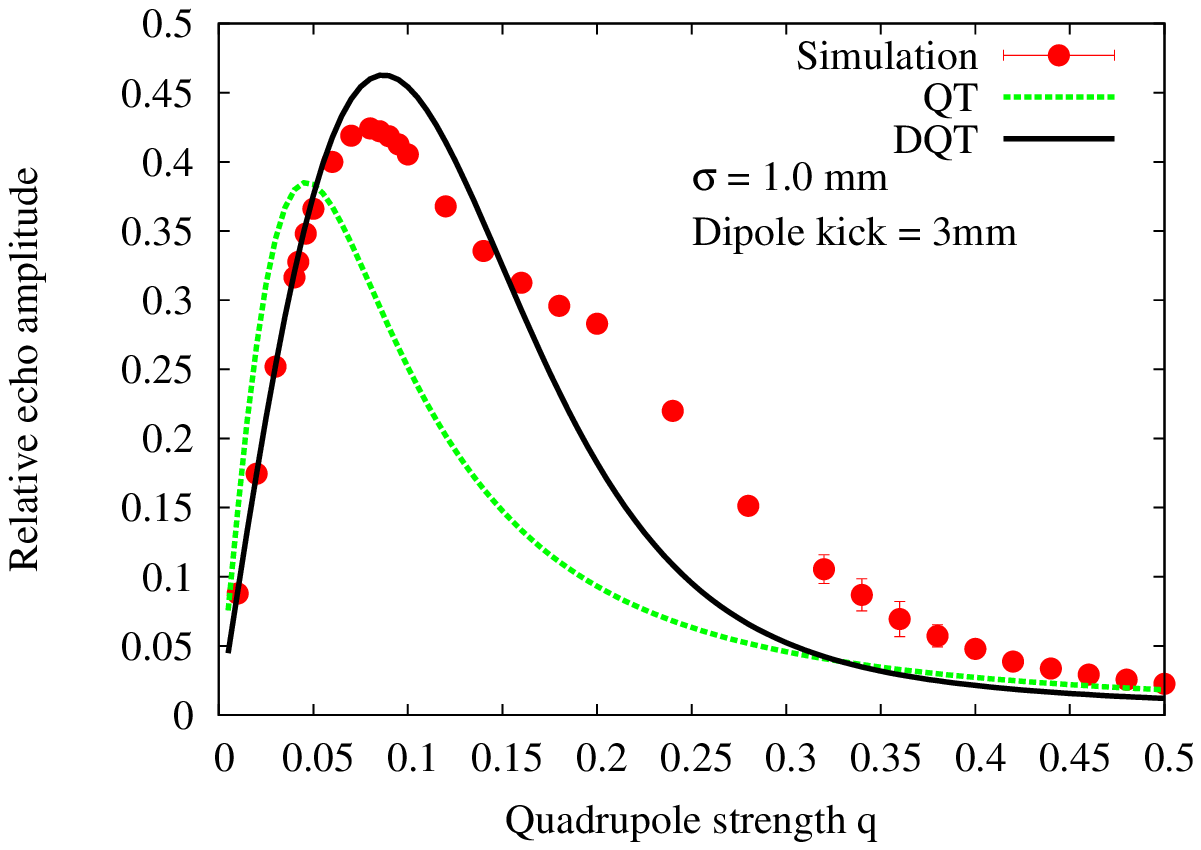}
\includegraphics[scale=0.55]{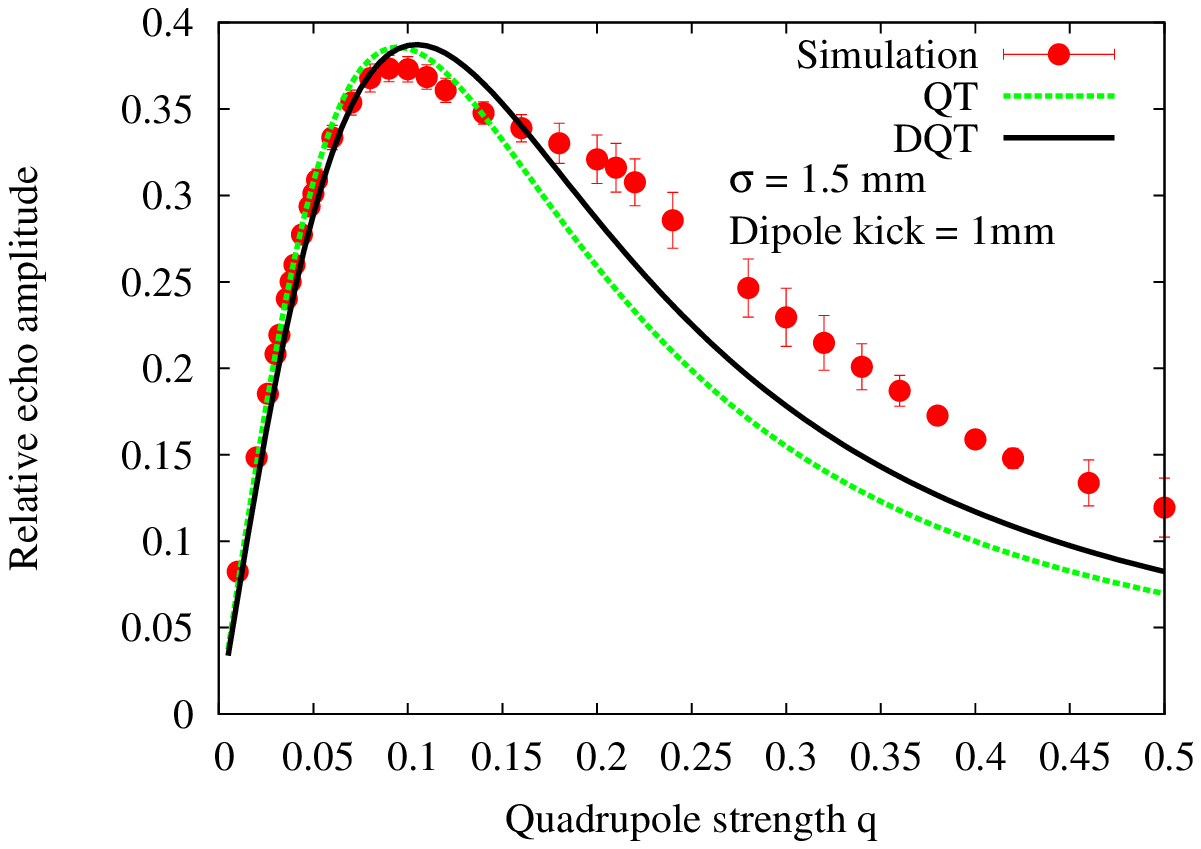}
\includegraphics[scale=0.55]{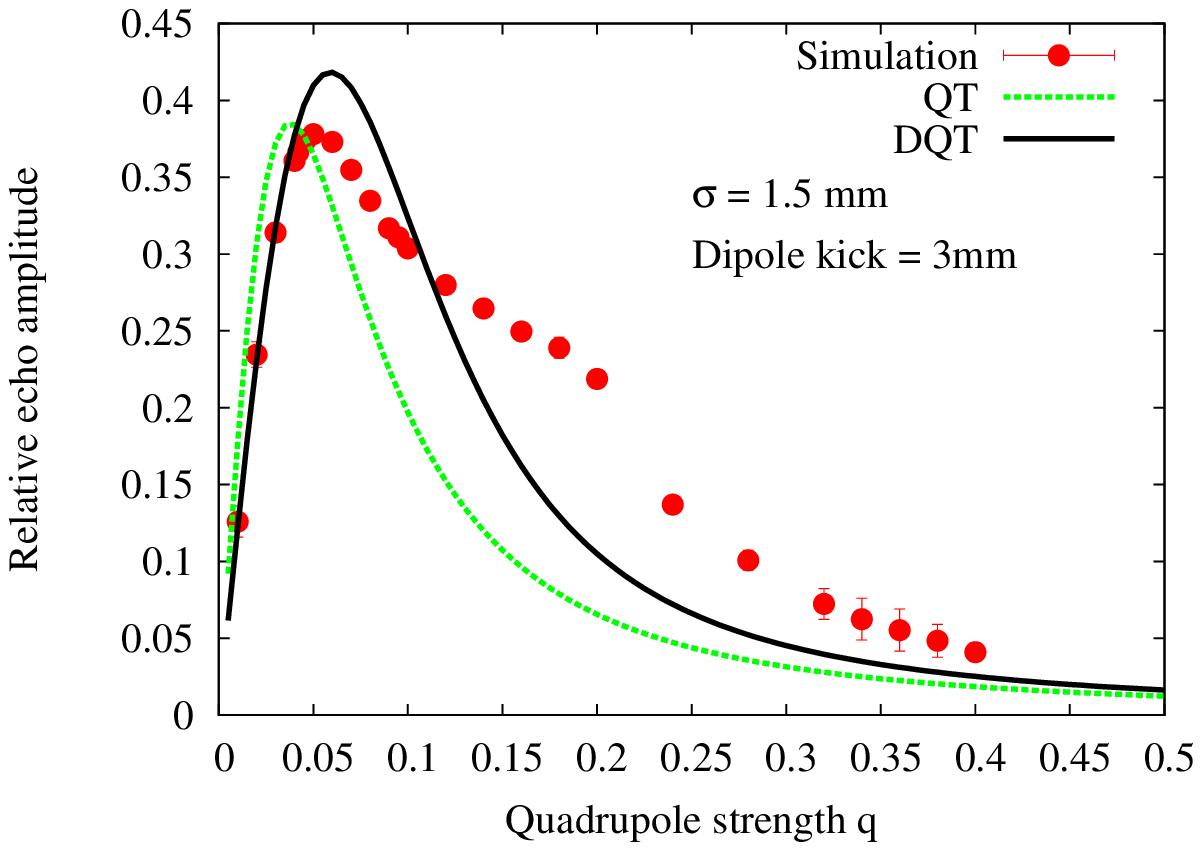}
\caption{The relative amplitude of the first echo as a function of the quadrupole strength parameter $q$. Simulations (red dots) 
are compared with QT, the nonlinear quadrupole theory (green curve) and DQT, the nonlinear dipole-quadrupole theory (black curve).
The initial emittances increase from top to bottom, at each emittance the left plot corresponds to a dipole kick= 1mm, the right
plot to a dipole kick = 3 mm.  }
\label{fig: amp_q_em}
\efig

\bfig
\centering
\includegraphics[scale=0.55]{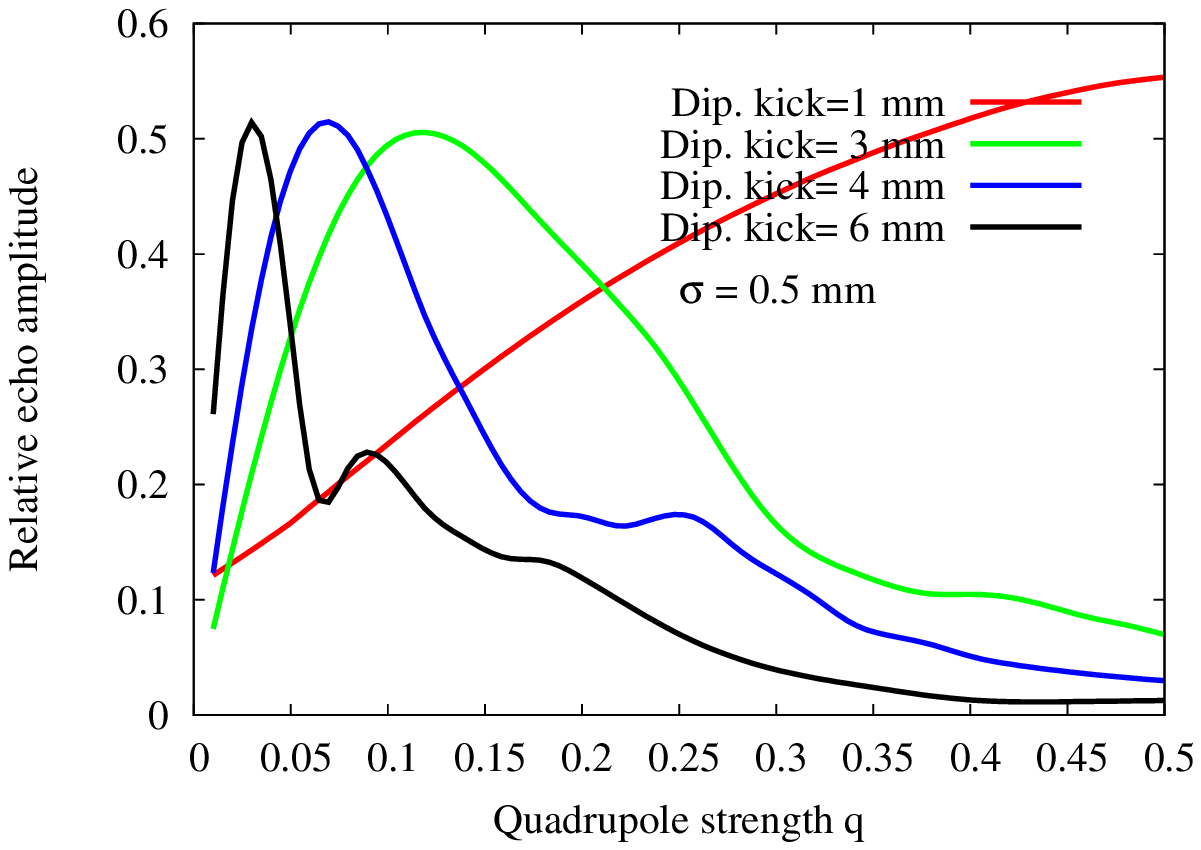}
\includegraphics[scale=0.55]{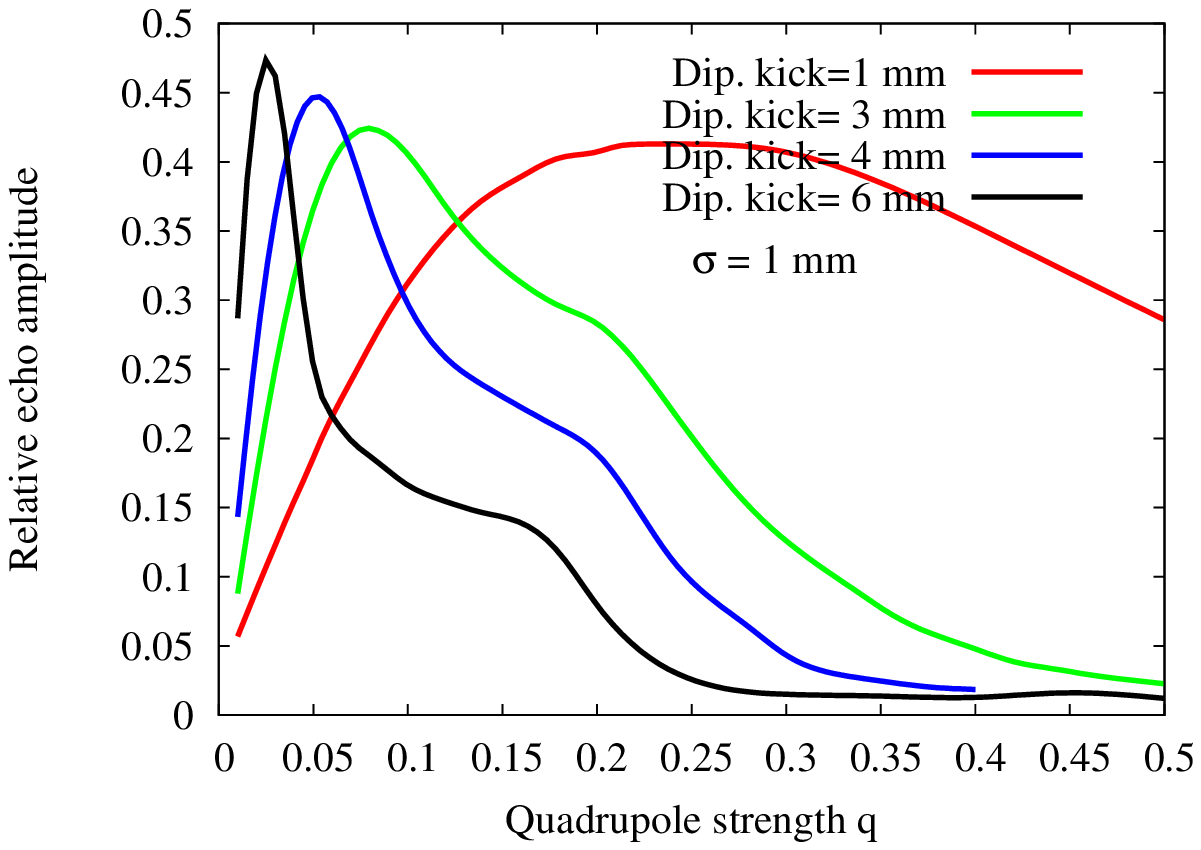}
\caption{Relative echo amplitude vs the quadrupole strength from simulations for different dipole kicks. 
Left: Initial beam size at the BPM $\sg=0.5$ mm; Right: $\sg=1$ mm. In both plots 
the maximum echo amplitude is not significantly affected by increasing the dipole kick, but the value of $q_{opt}$ changes
significantly.}
\label{fig: amp_q_alldip}
\efig
The plots in Fig.~\ref{fig: amp_q_alldip} show simulation results for the echo amplitude variation with $q$ over a large range of
dipole kicks. The left plot at the smaller initial emittance shows that 
at the smallest kick of 1 mm, the echo ampltude increases nearly linearly with $q$
and $A_{max}$ reaches a maximum value of about 0.55. 
As the dipole kick increases, the optimum quadrupole strength decreases, but there is little change in  $A_{max}$. 
The right plot in
Fig.~\ref{fig: amp_q_alldip} shows results at a larger initial emittance. The plots show similar behavior except that
the linear response is valid over a smaller range in $q$.
These simulation results confirm the results from theory that larger dipole kicks do not 
significantly impact the  amplitude of the first echo. 

Figure \ref{fig: qopt_emit} shows simulation results for the variation of $q_{opt}$ with the emittance for dipole kicks of 1mm and 
3mm. At the larger dipole kick, $q_{opt}$ values are about an order of magnitude smaller over most of this range of emittances, 
except at the largest emittances. 
\bfig
\centering
\includegraphics[scale=0.55]{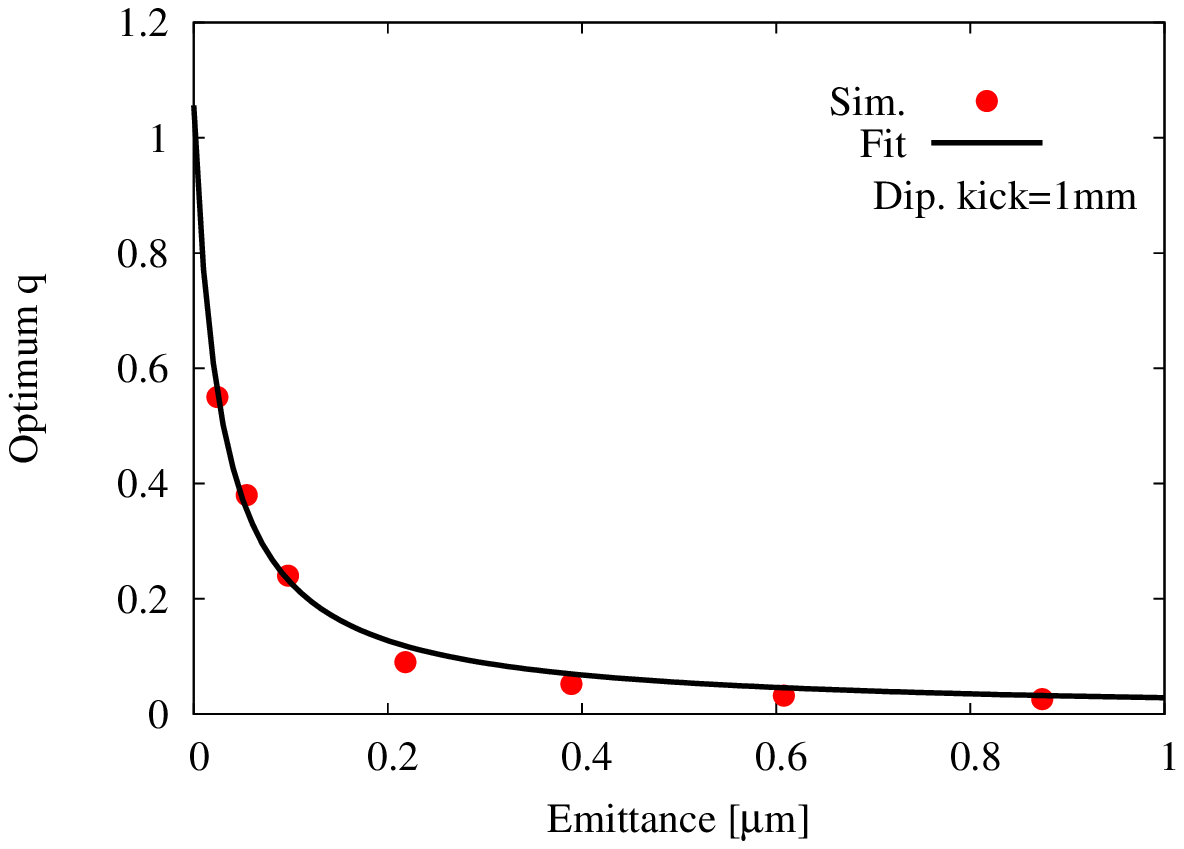}
\includegraphics[scale=0.55]{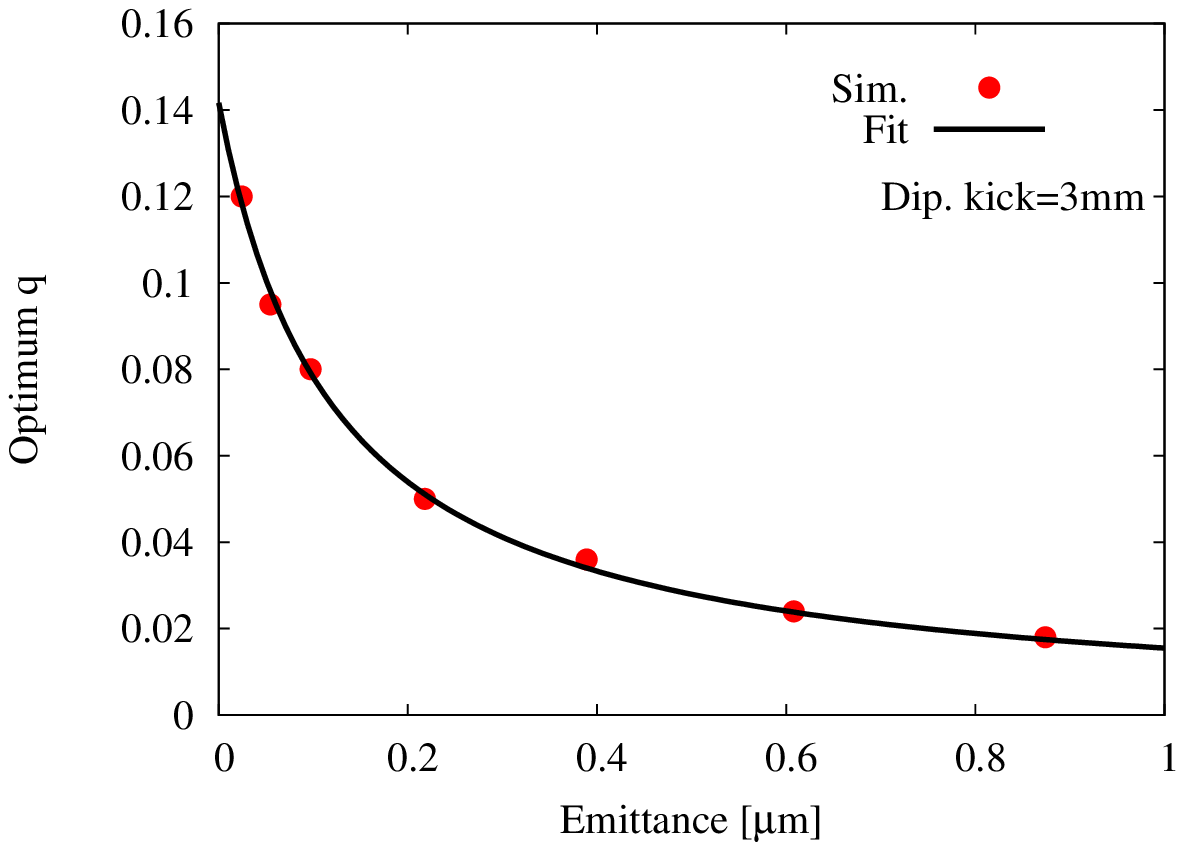}
\caption{Optimum quadrupole strength as a function of the initial emittance for two initial dipole kicks: simulations 
(red dots) compared with fits to the form in Eq.~(\ref{eq: qopt_fit}).}
\label{fig: qopt_emit}
\efig
The plots in Fig.~\ref{fig: qopt_emit} also show a fit to a function  
\beq
q_{opt}(\eps_0) = \fr{a_q}{\eps_0 + b_q}      \label{eq: qopt_fit}
\eeq
 where $(a_q, b_q)$ are fit parameters. This fit function  models the variation of $q_{opt}$ quite well in all the cases studied. 

\bfig
\centering
\includegraphics[scale=0.55]{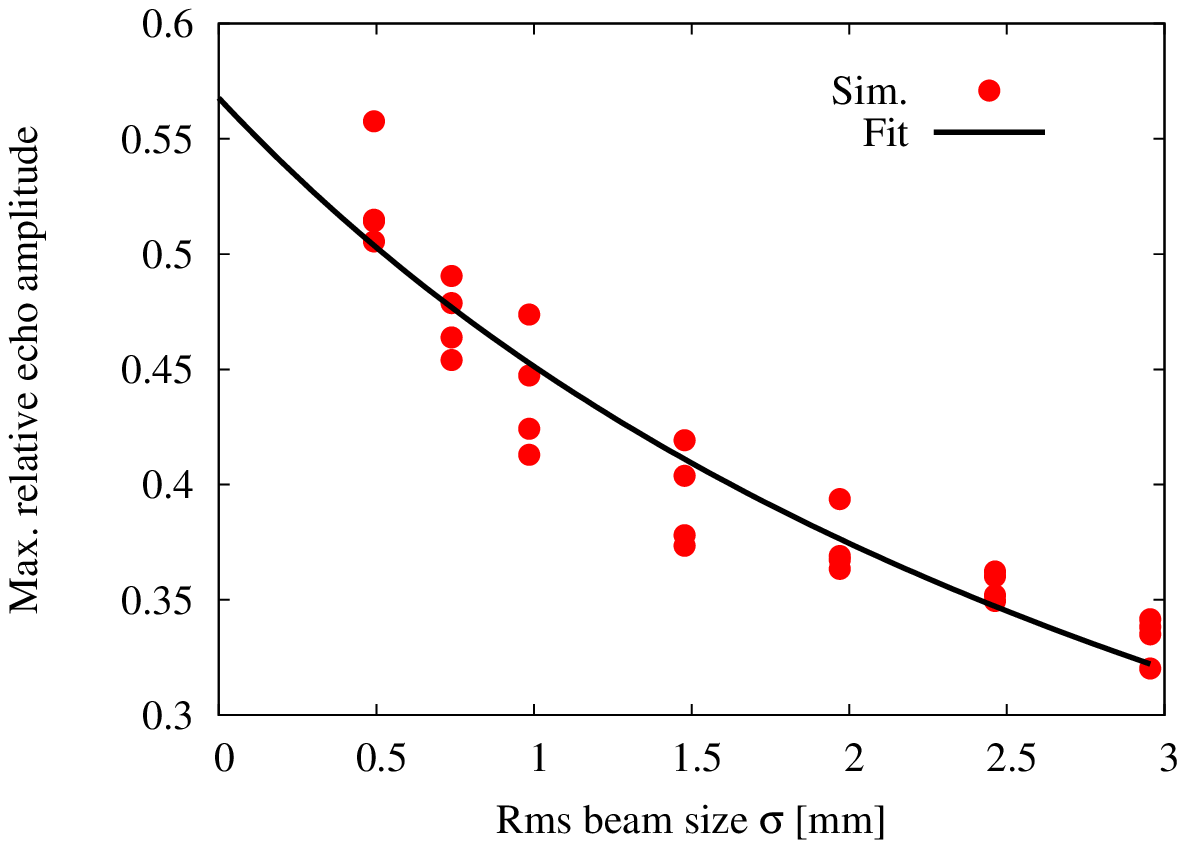}
\includegraphics[scale=0.55]{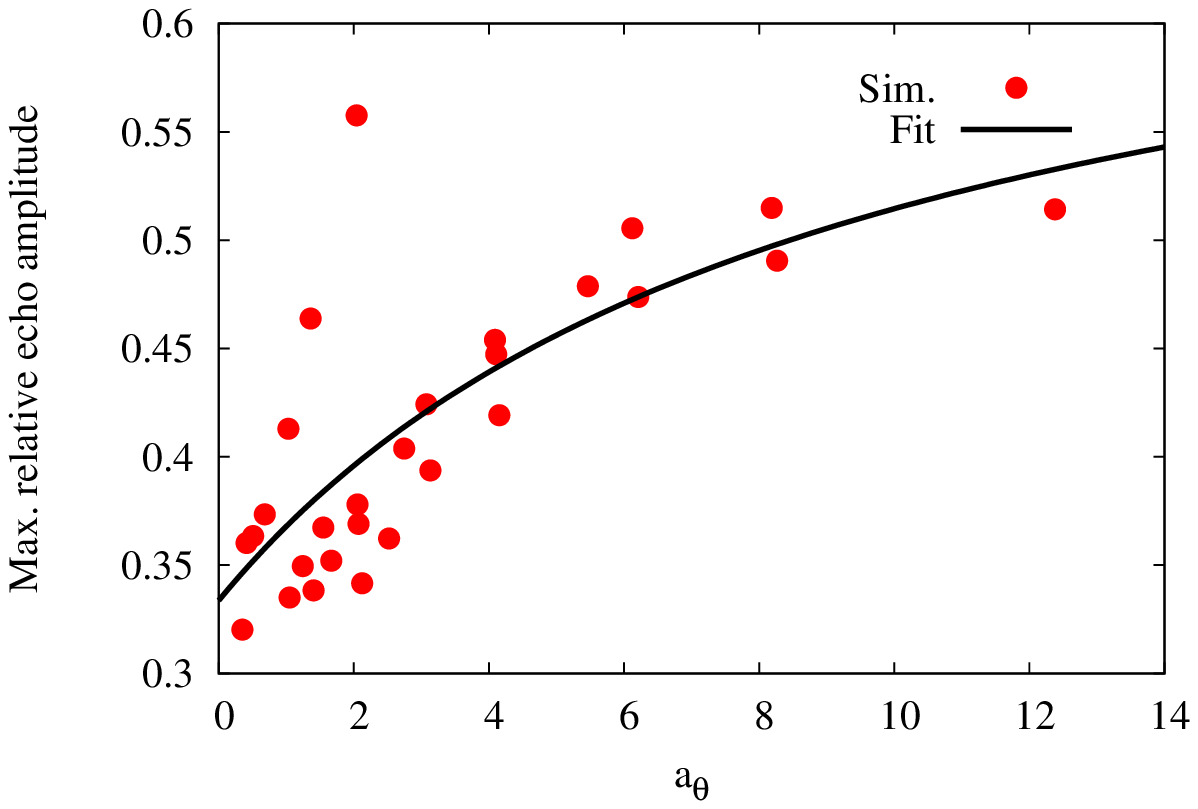}
\caption{Maximum relative echo amplitude $A_{max}$ as a function of the beam size (left plot) and the relative dipole kick (right plot)}
\label{fig: maxecho_sigx_dip}
\efig
The DQT theory in Section \ref{sec: NL_dip} had shown that $A_{max}$ is determined by  the emittance and the relative dipole
kick $a_{\theta}$, when the delay $\tau$ and detuning are kept constant. 
The left plot in Fig.~\ref{fig: maxecho_sigx_dip} shows simulation results for $A_{max}$ as a function of the initial beam size $\sg_0$, while the right plot shows $A_{max}$ as a function of  $a_{\theta}$. We find that $A_{max}$ as a function of $\sg_0$ is 
best fit by a functional form
\beq
A_{max}(\sg_0) = \frac{a}{\sg_0 + b}
\label{eq: Amp_max_fit_sig}
\eeq
where $(a, b)$ are fit parameters. It follows that  the maximum possible relative echo amplitude  at 
vanishingly small emittance is  $A_{asymp, \sg} \equiv A_{max}(\sg\rarw 0) = a/b$.  On the other hand,
$A_{max}$ as a function of $a_{\theta}$ is well fit by a rational function of the form
\beq
A_{max}(a_{\theta}) = \fr{p a_{\theta} + q}{ a_{\theta} + s}
\label{eq: Amp_max_fit_dip}
\eeq
where $(p, q, s)$ are fit parameters.
This function predicts that at very large $a_{\theta}$, the asymptotic value is given by
 $A_{asymp, a_{\theta}} \equiv A_{max}(a_{\theta}\rarw \infty) = p$. 
The plots in Fig.~\ref{fig: maxecho_sigx_dip} show the best fits to these functional forms. While there are the same 
number of simulation points in both plots, the scatter of points around the best fit is much smaller in the left
plot. The asymptotic values  predicted by the two fits are $A_{asymp, \sg}= 0.57$ 
and $A_{asymp, a_{\theta}}= 0.68$. 
The value of $A_{asymp, \sg}$ is much closer to the largest value seen in the simulations while reaching the value of
$A_{asymp, a_{\theta}}$ may require unrealistically large values of the dipole kick. 
These results again confirm that the maximum echo is largely determined by the initial beam  emittance; the variation with
dipole kick at a given emittance is within 15\% over the dipole kicks shown. 
A test beam with the smallest feasible emittance and modest dipole kick $a_{\theta} \sim 1$ may suffice to maximize the relative
echo amplitude. While the absolute echo amplitudes increase with the dipole kick, 
amplitudes  $\ge 0.1$ mm can be measured accurately when BPM resolutions are of the order of tens of 
microns. However, the advantage of a larger dipole kick, as seen 
in Fig.~\ref{fig: qopt_emit} is that the optimum quadrupole  strength is 
smaller, by up to an order of magnitude depending on the emittance. 
 In general, smaller dipole kicks are to be preferred
since they are less likely to lead to beam loss. In practice, generating the largest amplitude echo may require
a compromise between the largest dipole kick tolerable and quadrupole kick strengths achievable. 
Studies of stimulated echoes (not discussed here) show that a single large quadrupole kick can be
replaced by a few lower strength quadrupole kicks, spaced apart in time depending on the tune.

% \clearpage

\subsection{Multiple Echoes}

Multiple echoes could be useful to observe for the information they may provide about the
 machine and beam, such as diffusion and nonlinearities. It is also possible to enhance the multiple echoes with
different sequences of quadrupole pulses (stimulated echoes), so it is of interest to quantify their amplitudes
with just the single quadrupole kick studied in this paper. They were also observed during the echo experiments
at RHIC \cite{Fischer_2005}.

Simulations with 10,000  turns  are sufficient to observe up to the third
echo (if it exists) when the delay $\tau$ between the dipole and quadrupole kicks is 1400 turns. 
We find that increasing the quadrupole strength influences only the first echo but has no influence on the later
echoes which do not exist at small dipole kicks. 
The plots in Fig.~\ref{fig: mechoes_dip1-6_em3} show the evolution of the centroid at constant emittance and constant quadrupole
kick but increasing dipole kick. In this case, only the first echo  is seen at 1mm kick,  the  second echo is visible at a 3mm
kick while at a 6mm kick, both the second and third echoes are observed, with comparable amplitudes. 
\bfig
\includegraphics[scale=0.35]{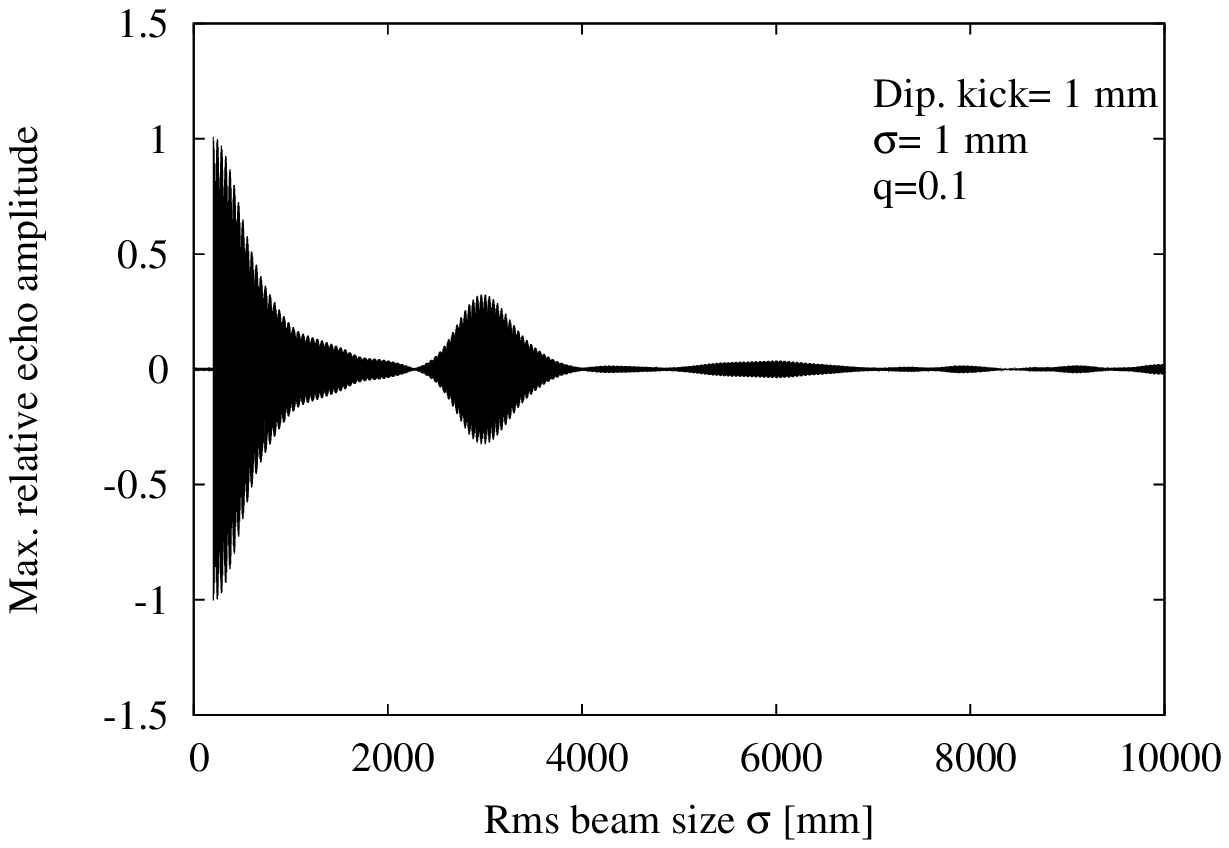}
\includegraphics[scale=0.35]{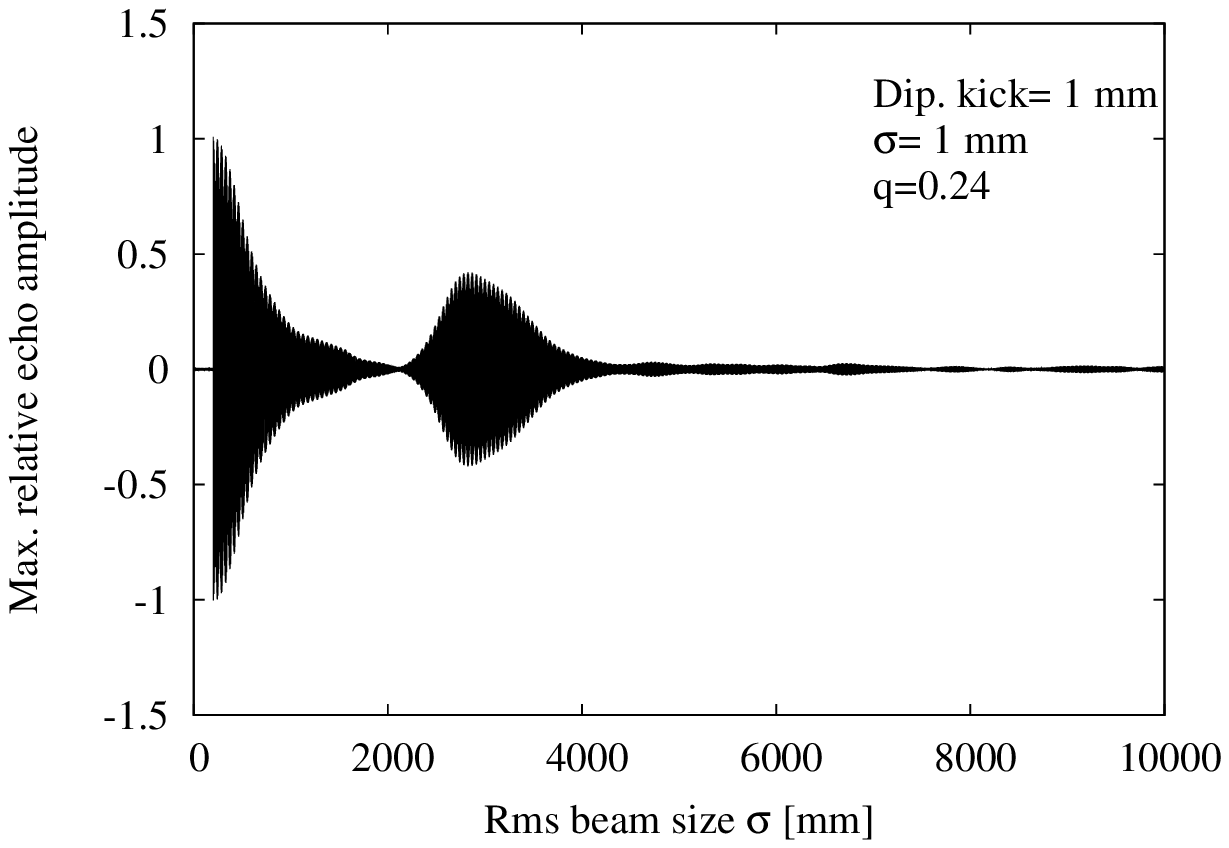}
\includegraphics[scale=0.35]{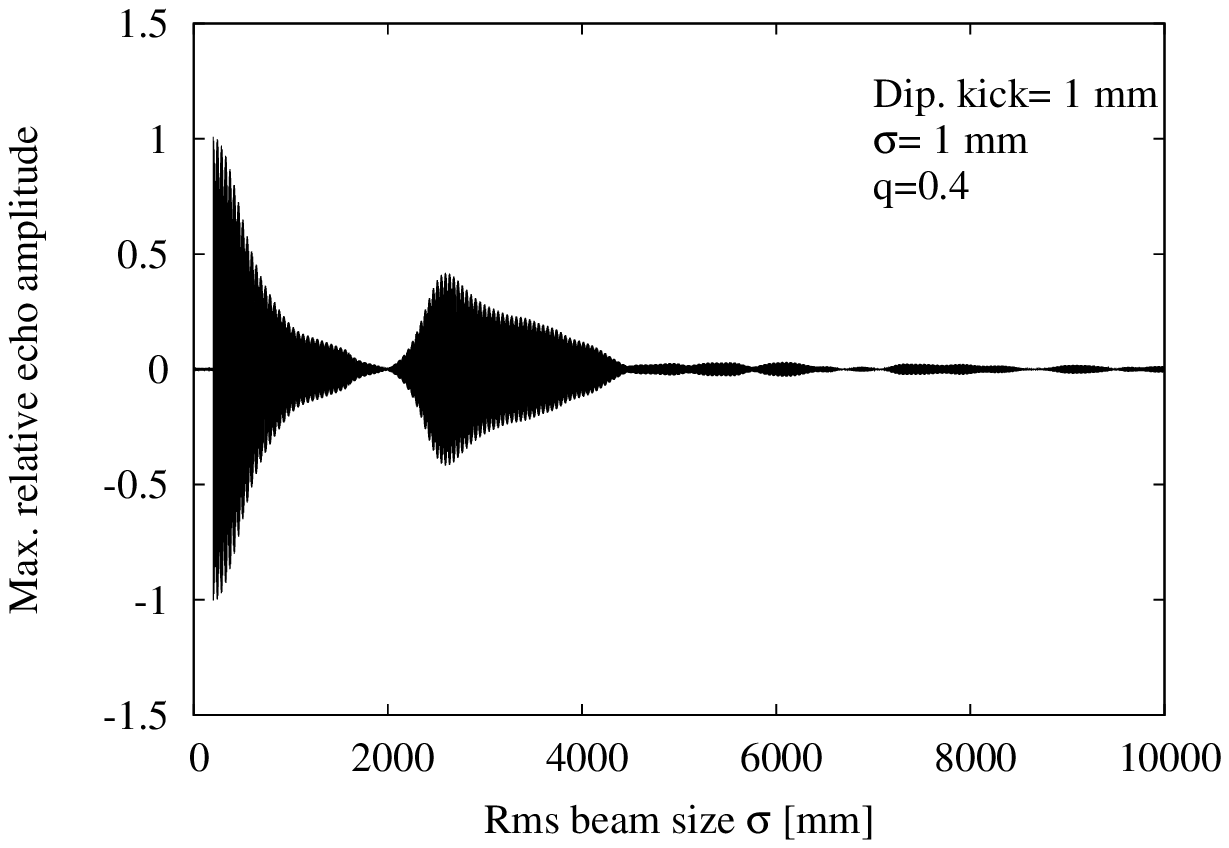}
\caption{Multiple echoes at constant emittance, quadrupole kick and increasing values of the dipole kick  strength, left to right. 
The second echo (centered at turn 5800) and third echo (centered at turn 8600) become visible at the larger dipole strengths.}
\label{fig: mechoes_dip1-6_em3}
\efig

Table \ref{table: echoes_1_2} compares the maximum relative amplitudes of the first and second echoes from 
theory and simulations at a constant emittance. The quadrupole strength was chosen such that it
led to the largest amplitude of the first echo. In most cases, theory and simulation results for the maximum amplitude are 
within 10\%. The only exception is the case with the second echo at  the smallest dipole kick of 1 mm; these amplitudes are very 
small in both theory and simulations. We also observe in the simulations that at $q=q_{opt}$ for the first echo, the second echo 
has started to bifurcate into two pulses, so values of $q < q_{opt}$ would be more suitable for the optimal second echo. 
\begin{table}
\bec
\btable{|c|c|c c|c c|} \hline
Dipole kick[mm] & Quad. strength $q$ & \multicolumn{2}{|c|}{1st echo amplitude} & \multicolumn{2}{|c|}{2nd echo  amplitude} \\
  &    &    Theory   & Simulation  &  Theory  & Simulation \\ \hline
1.0 &  0.24 &  0.39  &  0.42  &  0.04  &  0.024 \\
3.0  & 0.08  &  0.45  &  0.42  & 0.15  &  0.13 \\
4.0  & 0.052 &  0.48  &   0.45 & 0.17  &  0.16 \\
6.0  & 0.026 &  0.49  &  0.47  & 0.17 &  0.18 \\ \hline
\etable
\eec
\caption{Maximum relative amplitudes of the first and second echoes from theory and simulations. The emittance is constant
at $\eps_0=9.7 \times 10^{-8}$ m or rms size $\sg_0= 1$ mm. For each dipole kick, the quadrupole strength
is chosen  from simulations that maximizes the first echo amplitude.}
\label{table: echoes_1_2}
\end{table}
Simulations also validate the theoretical result from DQT that the amplitudes of the second and later echoes increase significantly 
with the dipole kick.

\setcounter{equation}{0}
\section{Spectral analysis of the echo pulse} \label{sec: spectrum}

The time dependent echo pulse shows that the amplitude is modulated at a frequency shifted from the betatron frequency. In the completely linear theory \cite{Chao} and Eq.~(\ref{eq: amp_linear}) in Section \ref{sec: NL_quad}, the time 
dependent pulse is $\lan x(t)\ran = \bt_K \theta Q A_F(t)$ where
\beqr
A_F(t) & = & \frac{1}{(1 + \xi(t)^2)^{3/2}}\sin [\Phi +  3\Theta(t)] , \;\;\;  \Phi = \om_{\bt}(t-2\tau), \;\;\;\; \Theta = {\rm Arctan}[ \xi(t)]
\eeqr
Since $\xi(t) = \om' \eps_0 (t - 2\tau)$, the lattice nonlinearity parameter $\om'$ can be retrieved from the frequency spectrum.
Taking the Fourier transform,
\beqrs
\tilde{A}_F(\om) & = & \int_{-\infty}^{\infty} dt \;  e^{i\om t} A_F(t)  = \frac{1}{2i}\int_{-\infty}^{\infty} dt \; e^{i\om t} \frac{1}{(1 + \xi^2)^{3/2}}
\left[e^{i(\Phi+3 \Theta)}-e^{-i(\Phi+3 \Theta)} \right] 
\eeqrs
The first term contributes to the negative frequency spectrum while the second contributes
to the positive frequency part. Considering the second term
\beqrs
\tilde{A}_F(\om > 0) & = &   -\frac{1}{2i} e^{2i \om_{\bt}\tau} \int_{-\infty}^{\infty} dt \; e^{i(\om - \om_{\bt})t} \frac{1}{(1 + \xi^2)^{3/2}}
e^{-3 i \Theta}
\eeqrs
This can be evaluated by a contour integration method, see Appendix B. 
The result for the echo spectrum as a function of frequency is
\beqr
\tilde{A}_F(\om) & = &  \left\{ \begin{array}{cc} -\frac{\pi}{6}\frac{e^{i2(\om - \om_{\bt})\tau}}{\mu\om_{rev}}
\dl^3 e^{-\dl}, \;\;\;\; &  \dl  \ge 0   \\
 0  , \;\;\;\; &   \dl < 0 \end{array} \right.   \\
\dl & \equiv &  \frac{\om - \om_{\bt}}{\mu\om_{rev}} = \fr{\nu - \nu_{\bt}}{\mu} 
\eeqr
where $\mu = \om'\eps/\om_{rev}$ is the tune shift at the rms beam size. 
This result shows first that the spectrum is non-zero only on one side of the nominal tune $\nu_{\bt}$: 
above $\nu_{\bt}$ if $\mu > 0$ or below $\nu_{\bt}$ if $\mu < 0$.
It also follows that the non-zero part of the spectrum has a peak at $\dl = 3$ or at a tune
given by
\beq
\nu_{peak} = \nu_{\bt} + 3\mu
\eeq

One measure of the width of the spectrum is the full width at half maximum (FWHM), which we find numerically to be 
$\dl_{FWHM} = 4.13$. Hence in tune space, the FWHM is
\beq
\Dl\nu_{FWHM} = 4.13\mu 
\label{eq: nu_fwhm}
\eeq
Thus both the tune of the echo pulse as well as the width of the echo spectrum are  related to the detuning. 
From the uncertainty relation for Fourier transforms $\Dl t \Dl \om \ge 1/2$ and using the FWHM for the echo pulse in time
\cite{Chao, Sen_2017},  $\Dl t_{FWHM} = 1.53/(\om_{rev}\mu)$, we expect that $\Dl\nu \ge \mu/3.06$. If we interpret the
FWHM as a measure of the uncertainty (although the rms spread is the usual measure), then Eq.~(\ref{eq: nu_fwhm}) satisfies the 
uncertainty relation.

The tune shift itself can be calculated simply from the time derivative of the phase $\Phi + 3\Theta$ and assuming that
${\rm Arctan}[\xi] \approx \xi = \om' \eps_0 (t - 2\tau) $ which is valid near the center of the echo at $t = 2\tau$. This yields
$\om \approx \om_{\bt} + 3 \om'\eps_0$, the same as the exact result. The additional advantage of the Fourier transform is 
that we also obtain the echo spectrum shape and width. 

The echo spectrum can also be calculated in the linear dipole kick and nonlinear quadrupole kick regime, when the time dependent echo pulse is
given by Eq.~(\ref{eq: pulse_nonlinq}). It has the same form as that in the completely linear regime, the time dependent phase shift from the 
betatron phase $\Phi_{\bt}$ is again $ 3\Theta(t)$ where now $\Theta$ is given by Eq.~(\ref{eq: Theta}). Using the same 
approximation of a small
argument of the ${\rm Arctan}$ function, we have for the angular betatron frequency  shift
\beq
\Dl\om = 3 \frac{d}{dt}\Theta \approx 3 \frac{d}{dt}[\frac{\xi }{1 - \xi^2 + Q^2}] \approx 3 \fr{ \om'\eps }{1 + Q^2}
\eeq
where we assumed $\xi \ll 1$ in the denominator and included the contribution of the dipole kick to the
emittance. Thus the nonlinearity of the quadrupole
kick will reduce the tune shift by a small amount from the linear regime, assuming $Q^2 < 1$.

\subsection{FFT of the echo pulse}

Here we use the simulation code to calculate the spectrum of the echo pulse and compare the results with the theory developed 
above. 
\bfig
\centering
\includegraphics[scale=0.55]{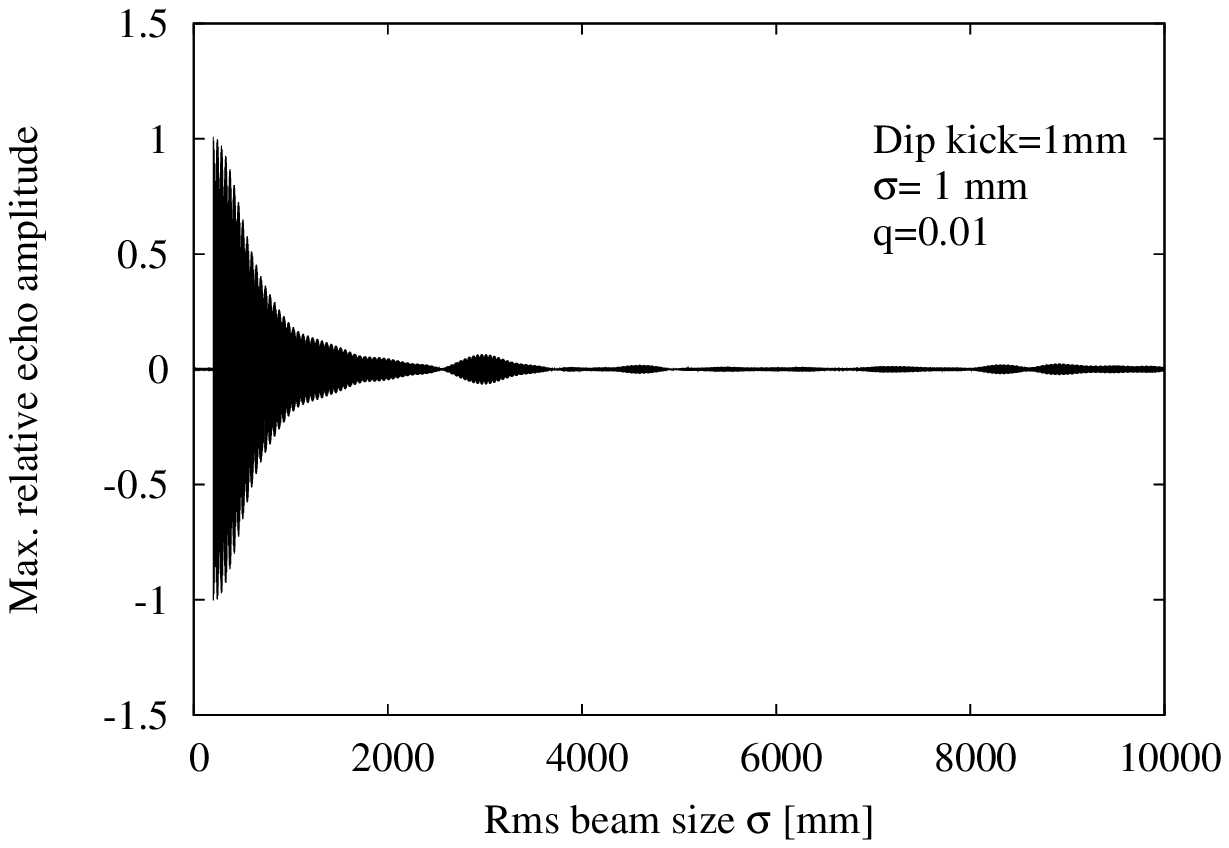}
\includegraphics[scale=0.55]{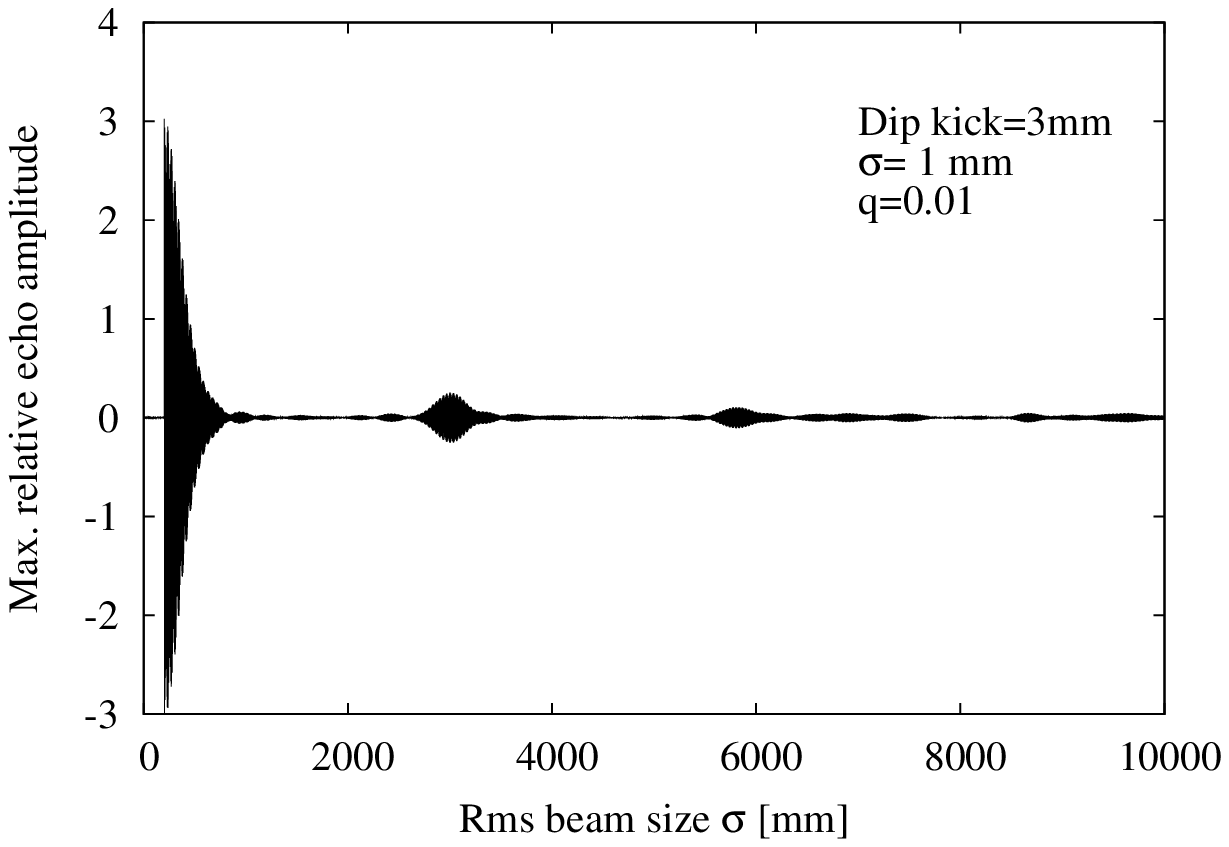}
\caption{Left: Tune shifts (without echoes) vs the emittance. The emittance was changed by varying dipole kicks.  Also shown is 
the straight line fit which yields the tune slope parameter $\nu'$. Right: Spectra without echo and with echoes. The spectrum 
without an echo
was obtained with $q=0$ while the echo spectra were obtained with the same dipole kick (1 mm), the same value of
$q=0.01$ and two initial emittances corresponding to $\sg_0=1.5$ mm  and $\sg_0=2$ mm. 
The  vertical dashed line shows the bare lattice tune.}
\label{fig: fft_slope_spectra}
\efig
One way of measuring the detuning parameter is to kick the beam to a range of amplitudes with varying dipole strengths. Each 
dipole
kick excites the beam to a different emittance allowing the betatron tune to be measured as a function of emittance. The left plot
in Figure \ref{fig: fft_slope_spectra} shows an example in our case. Here the quadrupole kick was set to zero so that no echoes
are excited and the initial emittance ($\sg_0=1$ mm) was kept constant. Dipole kicks over a range of 0.5-10 mm were used to vary
the final  emittance. Using the centroid data around the time of the echo formation for the FFT analysis
ensures that the beam has decohered to its asymptotic emittance. As expected the tune shifts in this plot lie on a straight line
and yield the tune slope as $\nu' = d\nu/d\eps = -3009$ m$^{-1}$. 
 The right plot shows spectra with and without echoes from an analysis of the centroid data using 1024 turns centered at the first
echo. The spectra with echoes are shown for two initial emittances and the same dipole kick of
1 mm. The beam is kicked to the same amplitude, but as the theory predicts, the negative detuning parameter
causes the echo spectrum to shift to the left and the spectrum widens with increasing
initial emittance. Table \ref{table: tunes_compare} shows a comparison of the simulated tune shifts and the theoretical value 
expected from the analysis above. 
\begin{table}
\bec
\btable{|c|c|c|} \hline
 Final emittance  & Theoretical $\Dl\nu$ & Simulated $\Dl\nu$ \\ 
 $\eps$ [$\mu$m]  & $\Dl\nu = 3\nu' \eps$   &   \\  \hline
  0.27  &  -0.0024   &  -0.0023 \\
 0.35  &  -0.0031  &  -0.0028 \\
  0.44  &  -0.0039   &  -0.0038 \\
  0.54  & -0.0049    &  -0.0043 \\
  0.65   & -0.0059   &  -0.0057 \\
\hline
\etable
\eec
\caption{Example of using the echo spectrum to measure the detuning, using a small amplitude dipole kick. All the echoes were 
generated with the same dipole kick of 1 mm and the same quadrupole kick $q=0.01$. 
The final un-normalized emittance is shown in the first column. In all cases,  the emittance increased by  $\Dl \eps=0.05$ $\mu$m. 
The second and third columns show the theoretical and simulated tune shifts respectively. The value of $\nu'=-3009$ /m 
was found using the simulation shown in the left plot of Fig.~\ref{fig: fft_slope_spectra}}.
\label{table: tunes_compare}
\end{table}
The prerequisites for using the echo spectrum to measure the detuning are that the initial beam decoherence must have a
negligibly small contribution to the echo, the echo pulse should be without distortions and obtained with small dipole and
quadrupole kicks so that the linear analysis is valid. Simulations of the echo spectrum at larger quadrupole strengths 
show that the echo tunes are not significantly affected, as expected from the analysis above. 
These results show that with some care, the echo spectrum can be used to measure the nonlinear detuning parameter 
without large amplitude dipole kicks.

\section{Conclusions}

In this paper we developed theories of one dimensional transverse beam echoes that are nonlinear in the dipole and quadrupole 
kick strength parameters with the goal of maximizing the echo amplitudes. Other relevant parameters are the initial beam 
emittance $\eps_0$, the freqency slope with emittance $\om'$  and the delay $\tau$ between the dipole and quadrupole kicks. 
The simpler theory (QT),  is linear in the dipole strength but nonlinear in the quadrupole strength $q$. 
This theory yields simple expressions for the optimum quadrupole strength $q_{opt}$ and the time dependent echo 
response. 
The optimum quadrupole strength is shown to decrease as the initial emittance and dipole kick strength increase.
This theory predicts that for emittances large enough that the decoherence time $\tau_D \ll \tau $, the maximum echo 
amplitude  relative to the dipole kick amplitude $A_{max} \approx 0.4$. 
Among the drawbacks of QT are that
it does not include ab initio the emittance growth due to the dipole kick, but has to be included as a correction. Nor does 
it predict the occurrence of  echoes at multiples of $2\tau$ beyond the first echo at $2\tau$. 
The second theory (DQT), which is nonlinear in both kicks, removes these drawbacks. The disadvantage is that
it results in more complicated expressions for the echo amplitude that require numerical integration. 
This theory predicts larger amplitude echoes than those with QT. It also 
 shows that increasing the dipole kick strength can reduce $q_{opt}$ by an order of magnitude but
has a minor influence on the relative amplitude of the first echo. 
However the amplitudes of later echoes at $4\tau, 6\tau,...$ increase significantly with the dipole kick. 

 One of the first observations from accompanying simulations was that  $\tau_D$ decreases with increasing either the initial 
emittance or dipole kick.  We found that at fixed detuning and delay, 
$A_{max}$ of the first echo  increases with smaller emittances but has a weak dependence on the 
dipole kick, in agreement with theory. In the limit of vanishing emittance $\lim_{\eps_0 \rarw 0} A_{max} = 0.57$
(see Fig.~\ref{fig: maxecho_sigx_dip}).  Both the QT and DQT are in good agreement with the
simulations for dipole kicks $\sim \sg_0$, the initial rms beam size. As a function of $q$, the 
echo amplitude from DQT was  in reasonable agreement with simulations for dipole kicks $\le 5\sg_0$. 
 For even larger larger dipole kicks, DQT yields acceptable results when $q \le q_{opt}$
but diverges from simulations for  $q \gg q_{opt}$. We attribute this to artifacts in the numerical integration which can be 
corrected.  Machine protection issues will forbid large dipole kicks, so in practice
DQT should be useful for estimating the echo amplitude. 
The simulations showed that the optimum quadrupole strength for higher order echoes changes with the echo order. 
Amplitudes of the later echoes increased with the dipole kick, again in accordance with the theory. 
The maximum amplitudes of the first and second echoes from theory and simulations agreed well, up to the largest dipole kick
(6$\sg$) tested. These results  suggest that a strategy for enhancing the echo signal would be to use
a pencil beam with reduced emittance, by scraping with collimators for example,  
(but with sufficient intensity to trigger the BPMs) and dipole kicks $\sim \sg_0$. The quadrupole strength should be scanned
in a range around $q_{opt}$ for the first echo to maximize its amplitude. If multiple echoes are not observed initially,  
increasing the dipole kick strength in incremental steps and rescanning around the appropriate $q_{opt}$ should reveal
their presence.

Spectral analysis of the echo pulse showed that the tune of the pulse is shifted from the bare betatron tune by
3$\mu$ where $\mu$ is the tune shift at the rms size. This was confirmed with simulations using small amplitude dipole 
and quadrupole kicks. This suggests that the 
echo pulses generated with small dipole kicks could be used to measure the detuning without the necessity 
of kicking the beam over a large range of amplitudes.

\vspace{2em}
\noi {\bf \large Acknowledgments} \newline
We thank the Lee Teng summer undergraduate  program at Fermilab for awarding an internship to Yuan Shen Li in 
2016. Fermilab is operated by the Fermi Research Alliance, LLC under U.S. Department of Energy contract No. DE-AC02-07CH11359.

%\clearpage

\appendix
\section*{Appendices}
%\addcontentsline{toc}{section}{Appendices}
\renewcommand{\thesubsection}{\Alph{subsection}}

\subsection{Appendix: Complete theory of nonlinear dipole and quadrupole kicks}
\setcounter{equation}{0}
\renewcommand{\theequation}{A.\arabic{equation}}

Here we  consider the complete distribution function (DF) following the dipole kick without the simplifying approximations 
made in Section \ref{sec: NL_dip}. Using the notation from this section and keeping terms to $O(q)$,
 the DF at time  $\tau$ after the dipole kick, 
\beqr
\psi_5(z,\phi, t) & = & \frac{1}{2\pi \eps_0} \exp[-\frac{\bt_K\theta^2}{2 \eps_0}]\exp\left\{- \frac{1}{\eps_0}
[z \eps_0 (1- q\sin 2\phi_{-\Dl \phi} ) \right. \nonumber \\
& &  \left.  + \bt_K\theta \sqrt{\frac{2\eps_0 z }{\bt}}(1 - \half q\sin 2\phi_{-\Dl \phi})  
\sin (\phi_{-\Dl\phi} - \tau \om(z) - q \cos^2\phi_{-\Dl\phi}  + Q z\sin 2\phi_{-\Dl \phi}) ] \right\} \nonumber \\
\mbox{} 
\eeqr
We  define dimensionless parameters 
\beqr
a_{\theta} & = &  \frac{\bt_K\theta}{\sg_0},   \;\;\; b_1  =  q, \;\;\; 
b_2   =    \sqrt{2} a_{\theta},  \;\;\;\; b_3  =  \frac{\sqrt{2}}{4} q a_{\theta} , \;\;\;\;\; b_i \ge 0 
\eeqr
We have the following ordering hierarchy assuming $q \ll 1, a_{\theta} \sim O(1)$
\[   b_2 > (b_1, b_3),  \;\;\;\; b_3 > b_1 \; {\rm if} \; a_{\theta} > 2 \sqrt{2} \]
In the theory developed in Section \ref{sec: NL_dip}, we had  kept only 
$b_2$ and dropped $b_1, b_3$. 

We have for the dipole moment
\beqr
\lan x(t) \ran & = & \frac{\sqrt{2\bt \eps_0}}{2\pi} \exp[-\frac{\bt_K\theta^2}{2 \eps_0}]\int dz \; \sqrt{z} \exp[- z] 
T_{\phi}(z) \\
T_{\phi}(z)  & \simeq &  {\rm Re} \left\{  \int d\phi e^{i\phi} \exp\left[
b_1z\sin (2\phi_{-\Dl \phi} ) - b_2\sqrt{z} \sin( \phi_{-\Dl\phi} -\half q - \tau \om + Q z \sin 2\phi_{-\Dl \phi} ) \right. \right. 
\nonumber  \\ 
&  &  + b_3\sqrt{z} \cos\left(\phi_{-\Dl \phi} + \half q + \tau\om - Q z \sin 2\phi_{-\Dl \phi}\right)   \nonumber  \\
& & \left.  \left.   - b_3\sqrt{z} \cos \left(3\phi_{-\Dl \phi} - \half q - \tau\om + Q z \sin 2\phi_{-\Dl \phi} \right) \right] \right\}
\label{eq: Tphi_preBessel}
\eeqr
where we used the approximation in Eq.(\ref{eq: approx_tauD}). 
Using the generating function  expansions for the modified Bessel functions, we have
\beqrs
T_{\phi}(z) & = &  {\rm Re} \left\{  \sum_{k_1}\sum_{k_2}\sum_{k_3}\sum_{k_4}  i^{k_1 + k_2} (-1)^{k_4}  
 I_{k_1}(b_1z) I_{k_2}(b_2\sqrt{z}) I_{k_3}( b_3\sqrt{z}) I_{k_4}(b_3\sqrt{z}) \right. \\
& &  \times \exp[i( k_1 2\Dl\phi - k_2(\Dl\phi + \tau\om+ q/2) - k_3(\Dl\phi -  \tau\om - q/2) \\
&  &  - k_4(3\Dl\phi +  \tau\om + q/2))] \\
& & \left.  \int d\phi  \exp\left[ i \left( [1 - 2 k_1 + k_2  + k_3 + 3 k_4] \phi + (k_2+k_4 - k_3)Q z\sin  2\phi_{-\Dl \phi}\right) \right]  \right\}
\eeqrs
We expand into a Bessel function, integrate over $\phi$, replace $k_2$ by $2k_1 - k_3 - 3k_4 - 2l - 1$,  and drop the sum 
over $k_2$. After simplifying the phase factor, the integrated term is
\beqrs
T_{\phi}(z) & = &   2\pi {\rm Re} \left\{  \sum_{k_1}\sum_{k_3}\sum_{k_4} \sum_l  i^{k_1 - k_3 + k_4  - 1} (-1)^{k_1 + k_4 + l}  
\exp[ i  -\half q([2(k_1 - k_3 - k_4 - l) - 1)] \right.  \\
& &  I_{k_1}(b_1z) I_{2k_1 - k_3 - 3 k_4 - 2l - 1}(b_2\sqrt{z}) I_{k_3}( b_3\sqrt{z}) I_{k_4}(b_3\sqrt{z})
J_l( [2(k_1 - k_3 -k_4 - l ) - 1]Q z ) \\
& & \left.  \times \exp\left( i\left[ \om( t - 2\tau(k_1 -  k_3 - k_4 - l)  ) \right]\right) \right\}
\eeqrs
Since the amplitude is locally maximum when the phase factor vanishes, the form above shows that echoes occur at 
close to the times $t$ when $ t - 2\tau(k_1 - k_3  - k_4 - l  ) = 0$. 
As expected, this predicts echoes only at  times close to multiples of 2$\tau$. 
We replace $ k_1 - k_3  - k_4 - l  = n$, which leads to
\beqrs
T_{\phi}(z) & = &   2\pi {\rm Im} \left\{  \sum_{k_1}\sum_{k_3}\sum_{k_4} \sum_n  i^{k_1 - k_3 + k_4} (-1)^{k_1 + k_4 + n}  
\exp[  - i  \half q(2n  - 1)] \right.  \\
& &  I_{k_1}(b_1z) I_{ k_3 - k_4 + 2n - 1}(b_2\sqrt{z}) I_{k_3}( b_3\sqrt{z}) I_{k_4}(b_3\sqrt{z})
J_{k_1 - k_3 - k_4 - n}( [ 2 n - 1]Q z ) \\
& & \left.  \times \exp\left( i\left[ \om( t - 2n \tau  \right]\right) \right\}
\eeqrs
Using the phase variables $\Phi_n, \xi_n$ defined earlier in Section \ref{sec: NL_dip}, we can write the
complete expression for the dipole moment as (after replacing $k_3 \rarw k_2, k_4 \rarw k_3$)
\beqr
\lan x(t) \ran & = & \sqrt{2\bt \eps_0} \exp[-\frac{\bt_K\theta^2}{2 \eps_0}] {\rm Im}\left\{ 
\int dz \; \sqrt{z} \exp[-\{ 1  - i \xi_n  \}z]  \right.  \nonumber \\
& &  \sum_n \sum_{k_1}\sum_{k_2}\sum_{k_3} i^{k_1 - k_2 + k_3} (-1)^{k_1 + k_3 + n} e^{i[ \Phi_n - \half q (2n  - 1) ]} \nonumber \\
& &  \left.   I_{k_1}(b_1z) I_{ k_2 - k_3 + 2n - 1}(b_2\sqrt{z}) I_{k_2}( b_3\sqrt{z}) I_{k_3}(b_3\sqrt{z})
J_{k_1 - k_2 - k_3 - n}( [ 2 n - 1]Qz) \right\}
\label{eq: echo_t_NLdip_all}
\eeqr
This is the most general form of the time dependent echo. 
To recover the approximate theory of Section \ref{sec: NL_dip}, we put $b_1 = 0 = b_3$. Since 
$I_0(0)=1, I_{m\ne 0}(0) = 0$, this requires $k_1 = 0  = k_2 = k_3$. Using $J_{-n}(z) = (-1)^n J_n(z)$, we
recover the same expression as in Eq.\ref{eq: echo_t_NLdip_1}). In the limiting case of no dipole kick, then
$b_2 = 0 = b_3$ and using the same Bessel function properties, we find that the dipole moment vanishes,
as it should. In the other limiting case of no quadrupole kick, we have $b_1 = b_3 = Q = 0$ and we have
a non-zero contribution only with $k_1 = k_2 = k_3 = n = 0$ and we have
\beqr
\lan x(t) \ran_{q=0} & = & \sqrt{2\bt \eps_0} \exp[-\frac{\bt_K\theta^2}{2 \eps_0}] {\rm Im}\left\{ 
e^{i \om_{\bt}t} \int dz \; \sqrt{z} \exp[-\{ 1 - i \om'\eps_0 t  \}z]  I_{1}(b_2\sqrt{z})  \right\} \nonumber \\
& = & \bt_K \theta {\rm Im}\left\{ \fr{e^{i \om_{\bt}t}}{(1 - i \om'\eps_0 t)^2}
\exp\left[\frac{\bt_K\theta^2}{2 \eps_0} \fr{i  \om'\eps_0 t}{(1 - i \om'\eps_0 t)}\right] \right\}
\eeqr
The last expression is the same as that derived by earlier authors \cite{Meller, Chao}.

Returning to the general case with non-zero dipole and quadrupole kicks, we extract the dominant terms
contributing to the first and second echoes at 2$\tau$ and 4$\tau$, 
by setting $n=1$ and $n= 2$ respectively in Eq.~(\ref{eq: echo_t_NLdip_all}),
\beqr
\lan x(2\tau) \ran & \simeq & 
\sqrt{2\bt \eps_0} \exp[-\frac{\bt_K\theta^2}{2 \eps_0}] {\rm Im}\left\{ 
\int dz \; \sqrt{z} \exp[-\{ 1  - i\xi_1 \}z] \right.  \nonumber \\
& &  \sum_{k_1=-N_1}^{N_1}\sum_{k_2=-N_2}^{N_2}\sum_{k_3=-N_3}^{N_3} i^{k_1 - k_2 + k_3} (-1)^{k_1 + k_3 +1} 
e^{i[ \Phi_1 - \half q]} 
\nonumber \\
& &  \left.   I_{k_1}(b_1z) I_{ k_2 - k_3 + 1}(b_2\sqrt{z}) I_{k_2}( b_3\sqrt{z}) I_{k_3}(b_3\sqrt{z})
J_{k_1 - k_2 - k_3 - 1}( Qz) \right\}  \label{eq: echo_2tau_all}  \\
\lan x(4\tau) \ran & \simeq & 
\sqrt{2\bt \eps_0} \exp[-\frac{\bt_K\theta^2}{2 \eps_0}] {\rm Im}\left\{ 
\int dz \; \sqrt{z} \exp[-\{ 1 - i\xi_2 \}z] \right.  \nonumber \\
& &   \sum_{k_1=-N_1}^{N_1}\sum_{k_2=-N_2}^{N_2}\sum_{k_3=-N_3}^{N_3} i^{k_1 - k_2 + k_3} (-1)^{k_1 + k_3 } 
e^{i[ \Phi_2 - \fr{3}{2} q]} \nonumber \\
& &  \left.   I_{k_1}(b_1z) I_{ k_2 - k_3 + 3}(b_2\sqrt{z}) I_{k_2}( b_3\sqrt{z}) I_{k_3}(b_3\sqrt{z})
J_{k_1 - k_2 - k_3 - 2}( 3 Qz) \right\}  \label{eq: echo_4tau_all}
\eeqr
Here the summations are written to indicate that a finite number of terms are calculated. 
From Eq.~(\ref{eq: echo_4tau_all}), it is easily checked that there is no contribution to the echo at $4\tau$
from terms linear in the dipole kick. This confirms the result in Section \ref{sec: NL_quad} where the
analysis to first order in the dipole kick did not reveal the presence of multiple echoes.

The convergence of the above expansions is rapid when the dipole kick parameter $a_{\theta}$ is sufficiently small. For
large $a_{\theta}\gg 1$, which can happen with either a large dipole kick or small emittance or both, the above expansions 
do not converge rapidly enough to be usable in some instances. A different approach would be to use the smallness of the
parameter $b_1 \ll 1$ to expand $\exp[b_1 z \sin[ 2(\phi-\Dl\phi)-q/2]]$ in Eq.~(\ref{eq: Tphi_preBessel}) into a power series in 
$b_1$ instead. A similar approach
had been used in \cite{Sen_2004} in calculating beam-beam tune shifts due to long-range interactions and was found to
converge rapidly. We will not investigate this method further here. For the comparisons with simulations, we use the equations
Eq.~(\ref{eq: echo_2tau_all}) and (\ref{eq: echo_4tau_all}) above when they do converge rapidly and in other cases, use the more
approximate version developed in Section \ref{sec: NL_dip}.

\subsection{Appendix: Echo Spectrum by Fourier transform}
\setcounter{equation}{0}
\renewcommand{\theequation}{B.\arabic{equation}}

Consider the Fourier amplitude from Section \ref{sec: spectrum}
\beq
\tilde{A}_F(\om > 0) = -\frac{1}{2i}
e^{ 2i \om_{\bt}\tau} \int dt e^{-i(\om - \om_{\bt})t} \frac{1}{(1 + \xi^2)^{3/2}}e^{-3i \Theta}
\equiv e^{2i\om_{\bt}\tau} I(\om ) \label{eq: App_A_AF}
\eeq
Using
\[
{\rm Arctan}[x] = \frac{i}{2}\ln[\frac{1 - i x}{1+ix}]
\]
and the definition of $ \Theta = {\rm Arctan}[ \xi(t)]$, we have
\[ \exp[-i3 \Theta(t)] = i \frac{(\xi + i)^3}{(1+\xi^2)^{3/2}} \]
Hence the integral reduces to
\beq
I(\om) = i \int_{-\infty}^{\infty} dt \; e^{-i(\om - \om_{\bt})t} \frac{1}{(\xi - i)^3}
 = \frac{i}{\mu\om_{rev}}e^{-i((\om - \om_{\bt})2\tau)}\int_{-\infty}^{\infty} d\xi \;
\frac{e^{i\dl \xi}}{(\xi - i)^3} \label{eq: App_A_Iom}
\eeq
where we defined $\dl = (\om - \om_{\bt})/(\mu\om_{rev})$ and $\mu\om_{rev} = \om'\eps$. Complexifying $\xi \rarw z$
we consider the contour integral $\oint dz \; e^{i\dl z}/(z- i)^3$ over a semi-circular
contour with the radius at infinity. If $\dl > 0$, then we consider the positive half plane and the 
integral vanishes over the arc leaving only the contribution over the real axis. The
integrand has third order poles at $z=i$, hence
\beqr
\int_{-\infty}^{\infty}dz \; \frac{e^{i\dl z}}{(z - i)^3} & = & 
\oint_C dz \; \frac{e^{i\dl z}}{(z - i)^3} =  2\pi i \times {\rm Residue}[\frac{e^{i\dl z}}{(z - i)^3}]_{z=i} \nonumber \\
 & = & \frac{\pi}{3}\dl^3 e^{-\dl},  \;\;\;\;\;\;  {\dl > 0}
\eeqr
On the other hand if $\dl < 0$, we consider the lower half plane where again the 
contribution from the arc vanishes. However the integrand is analytic over the lower half
plane, hence the contour integral vanishes. Thus we have
\beq
\int_{-\infty}^{\infty}dz \; \frac{e^{i\dl z}}{(z - i)^3} = 0,  \;\;\;\;\;\;  {\dl < 0}
\eeq
Hence the Fourier integral for positive frequencies, after combining Eqs.  (\ref{eq: App_A_AF}), (\ref{eq: App_A_Iom}) and the 
above contour integrations, is
\beqr
\tilde{A}_F(\om > 0) & = &  - \frac{\pi}{6\mu\om_{rev}}
e^{-i(\om - 2\om_{\bt})2\tau} \dl^3 e^{-\dl} ,  \;\;\;\;\;\;  {\dl \ge 0} \nonumber \\
& = & 0 ,  \;\;\;\;\;\;  {\dl < 0}
\eeqr
The echo spectrum is determined by the Fourier amplitude 
$|\tilde{A}_F(\om)| = (\pi/(6\mu\om_{rev})\dl^3 e^{-\dl}$ for $\om \ge \om_{\bt}$ and 
vanishes for $\om < \om_{\bt}$, assuming $\mu > 0$ while the converse is true if $\mu < 0$.


\begin{thebibliography}{}
\bibitem{Hahn}E.I. Hahn, {\em Spin Echoes}, Phy. Rev. {\bf 80}, 580 (1950)
\bibitem{Dale}B.M. Dale, M.A. Brown and R.C. Semelka,{\em MRI: Basic Principles and Applications}, Wiley Blackwell (2015)
\bibitem{Kurnit}N.A.~Kurnit, I.D.~Abella and S.R.~Hartmann, {\em Observation of a Photon Echo}, Phys. Rev. Lett. {\bf 13}, 567
 (1964)
\bibitem{Gould}R.W. Gould, T.M. O'Neil and J.H. Malmberg, {\em }, {\em Plasma Wave Echo}, Phys. Rev. Lett. {\bf 19}, 219
 (1967)
\bibitem{Malmberg}J.H. Malmberg, C.B. Wharton, R.W. Gould and T.M. O'Neil,{\em Observation of Plasma Wave Echoes},
Phys. Fluids, {\bf 11}, 1147 (1968)
\bibitem{Andersen}M.F. Andersen, A.Kaplan and N. Davidson,{\em Echo Spectroscopy and Quantum Stability of Trapped Atoms},
Phys. Rev. Lett., {\bf 90}, 023001 (2003)
\bibitem{Yu}J.H. Yu, C.F. Driscoll and T.M. O'Neil, {\em Phase mixing and echoes in a pure electron plasma}, Phys. Plasmas,
{\bf 12}, 055701 (2005)
\bibitem{Karras}G. Karras, E. Hertz, F. Billard, B. Lavorel, J.-M. Hartmann, O. Faucher, Erez Gershnabel, Yehiam Prior, and Ilya Sh. Averbukh, {\em Orientation and Alignment Echoes}, Phys. Rev. Lett. {\bf 114}, 153601 (2015)
\bibitem{Stupakov} G.V.  Stupakov, Preprint, SSCL-579 (1992)
\bibitem{Stup_Kauf}G. V. Stupakov and S.K. Kaufmann, Preprint SSCL-587 (1992)
\bibitem{Fermi_AA}L.K.~Spentzouris,J-F.~Ostiguy, P.L.~Colestock, {\em Measurement of Diffusion Rates in High Energy 
Synchrotrons using Longitudinal Beam Echoes}, Phys. Rev. Lett., {\bf 76}, 620 (1996)
\bibitem{CERN_SPS}O.~Bruning, T. ~Linnecar, F. ~Ruggiero, W. ~Scandale, E.~ Shaposhnikova, D. ~Stellfeld, 
{\em Beam Echoes in the CERN SPS}, Proceedings of PAC97, 1816 (1997)
\bibitem{Arduini}G. Arduini, F. Ruggiero, F. Zimmermann, M. Zorzano-Mier, Preprint CERN-SL-Note-2000-048-MD (2000)
\bibitem{Fischer_2005}W.~Fischer, T.~Satogata and R.~Tomas, {\em Measurement of  Transverse Echoes in RHIC}, 
Proceedings of PAC2005, 1955 (2005)
\bibitem{Sen_2017}T.~Sen and W.~Fischer, {\em Diffusion Measurement from Observed Transverse Beam Echoes}, Phys. Rev. AB 
{\bf 20}, 011001 (2017)
\bibitem{Stancari}G.~Stancari, {\em Measurement of Beam Halo Diffusion and Population Density in the Tevatron and in the 
Large Hadron Collider},  Proceedings of HB2014, 294 (2014).
\bibitem{Chao}A.W.~Chao, Lecture Notes at www.slac.stanford.edu/$\sim$achao/lecturenotes.html
\bibitem{Meller}R. E.~Meller, A.W.~Chao, J.M.~Petersen, S.G.~Peggs and M.~Furman, SSC-N-360 (1987)
\bibitem{Ed_Sy}D.~Edwards and M.J.~Syphers, {\em An Introduction to the Physics of High Energy Accelerators}, Wiley
\bibitem{Grad_Ryz}I.S.~Gradshteyn and I.M.~Ryzhik,{\em Table of Integrals, Series and Products}, Academic Press (1983)
\bibitem{Abram_Steg}M.~Abramowitz and I. A.~Stegun,{\em Handbook of Mathematical Functions}, Dover Publications
\bibitem{Lucas}S.~K.~Lucas and H.~K.~Stone,{\em Evaluating infinite integrals involving Bessel function integrands of arbitrary order},
J.Comp.App.Math.,{\bf 64}, 217 (1995)
\bibitem{Sen_2004} T.~Sen, B.~Erdelyi, M.~Xiao and V.~Boocha, {\em Beam-beam effects at the Fermilab Tevatron: Theory},
  Phys. Rev. AB, {\bf 7}, 041001 (2004)
\end{thebibliography}
\end{document}